\documentclass[aps,twocolumn,showpacs,preprintnumbers,nofootinbib,prd,superscriptaddress,groupedaddress,10pt]{revtex4-1}

\makeatletter
\def\l@subsubsection#1#2{}
\def\l@subsubsubsection#1#2{}
\makeatother

\usepackage{subfigure}
\usepackage{graphicx,amssymb,amsmath,amsthm,amsfonts,epsfig,epsf}
\usepackage[usenames]{color}
\usepackage{epstopdf}
\definecolor{darkred}{rgb}{0.5,0,0}
\usepackage{latexsym}
\usepackage{array}
\usepackage{afterpage}
\usepackage{bm}
\usepackage{dcolumn}
\usepackage[utf8]{inputenc}
\usepackage{rotating}
\usepackage{longtable}

\usepackage{enumerate}
\usepackage{tensor,multirow}
\usepackage{url}
\usepackage{placeins}
\usepackage[linktocpage]{hyperref}
\usepackage{float}
\usepackage{mwe}
\usepackage[utf8]{inputenc}
\def\be{\begin{eqnarray}}
\def\ee{\end{eqnarray}}
\usepackage{soul}

\begin{document}

\title{Accelerating black holes: quasinormal modes and late-time tails}

\author{Kyriakos Destounis}
\email{kyriakos.destounis@uni-tuebingen.de}
\affiliation{Theoretical Astrophysics, IAAT, University of Tübingen, Auf der Morgenstelle 14, 
72076 Tübingen, Germany}

\author{Rodrigo D. B. Fontana}

\affiliation{Universidade Federal da Fronteira Sul, Campus Chapec\'o-SC
Rodovia SC 484 - Km 02, CEP 89815-899, Brasil}

\author{Filipe C. Mena}

\affiliation{Centro de An\'alise Matem\'atica, Geometria e Sistemas Din\^amicos, Instituto Superior T\'ecnico, Universidade de Lisboa, Avenida Rovisco Pais 1, 1049-001 Lisboa, Portugal}
\affiliation{Centro de Matem\'atica, Universidade do Minho, 4710-057 Braga, Portugal}

\date{\today}

\begin{abstract}
\noindent{ 
Black holes found in binaries move at very high velocities relative to our own reference frame and can accelerate due to the emission of gravitational radiation. Here, we investigate the numerical stability and late-time behavior of linear scalar perturbations in accelerating black holes described by the $C-$metric. We identify a family of quasinormal modes associated with the photon surface and a brand new family of purely imaginary modes associated with the boost parameter of the accelerating black hole spacetime. When the accelerating black hole is charged, we find a third family of modes which dominates the ringdown waveform near extremality. Our frequency and time domain analysis indicate that such spacetimes are stable under scalar fluctuations, while the late-time behavior follows an exponential decay law, dominated by quasinormal modes. This result is in contrast with the common belief that such perturbations, for black holes without a cosmological constant, always have a power-law cutoff. In this sense, our results suggest that the asymptotic structure of black hole backgrounds does not always dictate how radiative fields behave at late times.}
\end{abstract}

\maketitle
\section{Introduction} 

A direct consequence of the existence of binaries of black holes (BHs) \cite{Barack:2018yly} is the emission of gravitational waves, as the objects inspiral and merge to form a final compact object. During this process an immense amount of energy is released with a distinctive waveform that can be calculated using General Relativity (GR). Such waveforms have recently been detected by LIGO and Virgo detectors \cite{PhysRevLett.116.061102,PhysRevLett.116.241103,PhysRevLett.118.221101,PhysRevLett.119.141101,Abbott_2017}, leading the way to a new realm of observational astronomy.

The gravitational waveform produced by BH binaries has three distinct stages, namely, the inspiral, merger and ringdown stages. At the inspiral stage, the BHs orbit around a common center of mass. As they lose energy through the emission of gravitational waves, their orbits accelerate and the individual objects appear to move at very high velocities with respect to our own reference frame. 

Moreover, the anisotropic nature of gravitational wave emission from BH binaries leads to the emission of linear momentum. This can result in a recoil of the final remnant, known as a BH kick, which may be strong enough to eject a supermassive BH from its host galaxy \cite{Merritt:2004xa,1989ComAp..14..165R}, leading to BH remnants traveling through intergalactic space. This effect has been studied in detail using various numerical techniques \cite{Brugmann:2007zj,Centrella:2010mx}, including its potential imprint on the gravitational waveform \cite{Gerosa:2016vip,CalderonBustillo:2018zuq}. 
 In fact, some key findings indicate that the merger of non-rotating BHs can produce kicks of $\sim 170 \,\text{km s}^{-1}$ \cite{Gonzalez:2006md}, while mergers of rapidly rotating BHs can lead to kick velocities as high as $\sim 5000\, \text{km s}^{-1}$ (even $\sim 15000\, \text{km s}^{-1}$ in ultrarelativistic encounters) \cite{Gonzalez:2007hi,Campanelli:2007cga,Lousto:2011kp,Sperhake:2010uv}. 

Therefore, an understanding of moving and accelerating BHs, as well as the interaction with their astrophysical environment, is a paramount ingredient to explore the enormous possibilities of such sources of radiation \cite{Bernard:2019nkv,Cardoso:2019dte}.

An important question, both from the mathematical and physical points of view, is whether BH solutions are stable against small perturbations. Indeed, the response of BHs to external perturbations has a long-lasting history (see e.g. \cite{Chandrasekhar:1985kt}). When a BH is perturbed slightly, it exhibits damped oscillations, which may potentially be described by quasinormal modes (QNMs) \cite{Kokkotas:1999bd,Berti:2009kk,Konoplya:2011qq}. It turns out that QNMs are crucial at the ringdown stage of a waveform, where the remnant object oscillates until it relaxes to a final stable object. Thus, QNMs not only describe the oscillatory frequency and decay rate of the perturbed BH, but also carry key information about the externally observable parameters characterizing the final stable object.

In this article, we investigate, for the first time, the response of neutral and charged accelerating black holes to scalar perturbations. We use the $C-$metric originally found by Weyl in 1917 \cite{Weyl}. This spacetime is an axially symmetric exact solution to the Einstein-Maxwell equations. Its physical interpretation and causal properties were first discussed in \cite{Kinnersley_1970}, where it was shown that the $C-$metric, and its charged version, describe a pair of causally separated BHs which accelerate uniformly in opposite directions under the influence of a cosmic string, represented by a conical singularity. Later on, it was shown \cite{Ernst_1976,Ernst_1978} that the conical singularity can be removed by appending an external electromagnetic field, leading to the Ernst spacetime.
The geometrical and radiative properties of the $C-$metric were investigated in \cite{Bicak_1968,Farhoosh,Bicak:1989knv,Bicak_1999,Pravda:2000vh}, while its asymptotic properties were analyzed in \cite{ashtekar1981}, revealing that this metric is asymptotically flat but not asymptotically Euclidean (in the sense that conformal infinity is global, i.e., admits spherical sections, though its generators are not complete \cite{ashtekar1981}). 

The $C-$metric has been generalized to include rotation, cosmological constant and a Newman-Unti-Tamburino (NUT) parameter in \cite{Plebanski:1976gy} and was further analyzed in many different contexts (see e.g. \cite{Emparan:1999fd,Emparan:1999wa,Podolsky:2002nk,Dias:2002mi,Dias:2004rz,Podolsky:2000pp,Dias:2003xp,Dias:2003st,Griffiths_2005,Bicak_2010,Anabalon:2018qfv,Anabalon:2018ydc,Gregory_2019,Appels_2016}). For example, in the context of quantum gravity, charged C-metrics have been used to describe the production of BH pairs in strong background fields \cite{Hawking-Horowitz-Ross}.

In particular, the $C-$metric has been seen as a reasonable candidate to describe boosted BHs, by considering the fact that the oppositely accelerating BHs are causally disconnected and the metric can be expressed in appropriate coordinates to only cover one of the ``moving" BHs (see e.g. \cite{Hawking:1997ia}). 

We take advantage of this property and study, numerically, neutral massless scalar field perturbations, which propagate on the fixed background. We calculate the corresponding QNMs and conclude that accelerating black holes are modally stable against such perturbations. 

We also demonstrate that the charged $C-$metric entails a broader spectrum of QNMs, that is, three distinct families of modes co-exist. 
The first family is associated with the photon surface where null particles are trapped in unstable orbits. The decay timescales of those QNMs exhibit an anomalous behavior depending on the boost parameter of the accelerating spacetime. In fact, their timescale increases or decreases with increasing magnetic quantum numbers, depending on whether the spacetime boost is small or large. The second family consists of purely imaginary modes, governed by the boost parameter of the BH and the existence of an acceleration horizon. This family has never been observed before, as far as we know. The charged $C-$metric possesses a third family of modes, which are also purely imaginary, and become dominant when the event and Cauchy horizons approach each other and the BH tends to extremality. 

Due to the asymptotically flat nature of such metric, one might expect that a power-law cutoff will suppress the quasinormal ringing phase \cite{Price_1972}. We show that this is not the case here, as the late time behavior of perturbations involves an exponential-law decay, dominated by the QNMs of the scalar field. This can be understood from the fact that the presence of the acceleration horizon leads to initial data for the scalar field wave equation which do not intersect infinity, and thus, the asymptotically flat region is causally disconnected from the region between the event and acceleration horizon considered here.

In what follows, we will use geometrized units such as $c=G=1$.
%%%%%%%%%%%%%%%%%%%%%%%%%%%%%%%%%%%%%%%%
\section{The charged $C-$metric}
\begin{figure}[t]
\includegraphics[scale=0.5]{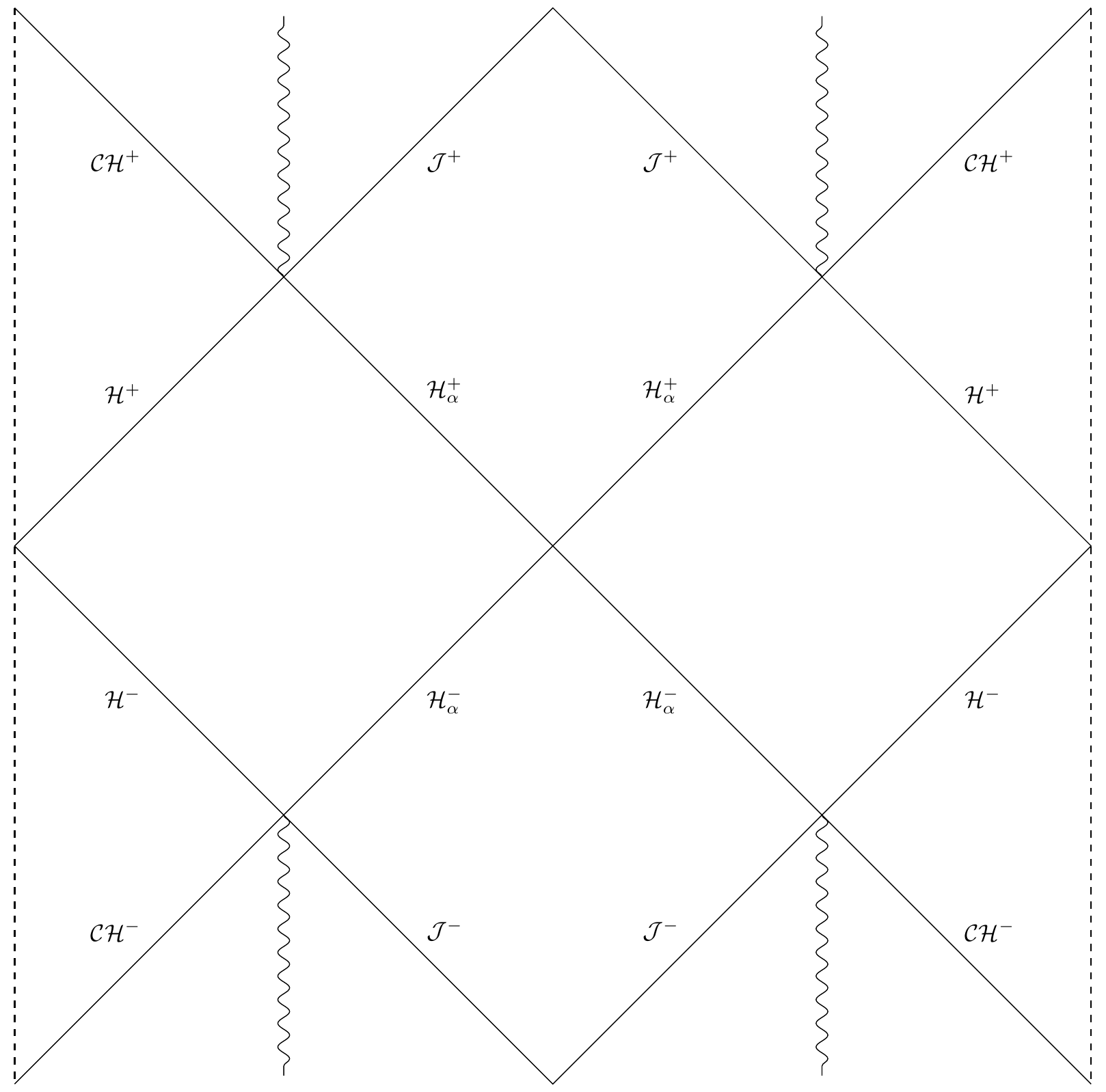}
\caption{The Penrose diagram, adapted from \cite{Hawking:1997ia}, of the charged $C-$metric, neglecting the axis $\theta=0$. The dashed lines are identified, while the wiggled lines correspond to curvature singularities at $r=0$. In turn, $\cal{H}^{\pm},\,\cal{H}_\alpha^{\pm},\,\cal{CH}^{\pm}$ and $\cal{J}^{\pm}$ correspond to the future $(+)$ and past $(-)$ event, acceleration, Cauchy horizon and null infinity.}
\label{Penrose}
\end{figure}

The $C-$metric describes a pair of causally separated BHs accelerating away from each other in opposite directions \cite{griffiths_podolsky_2009}. This metric is a generalization of the Schwarzschild solution which, besides the BH mass, includes an additional parameter related to the BHs acceleration. The charged version of the $C-$metric includes an electric charge parameter, related to an electromagnetic field, and the metric that covers one of the charged BHs can be written in spherical-type coordinates as \cite{Griffiths:2006tk,griffiths_podolsky_2009}
\begin{align}
\label{Cmetric}
\nonumber
ds^2=\frac{1}{\left(1-\alpha r \cos\theta\right)^2}\left(-f(r)dt^2+\frac{dr^2}{f(r)}\right.\\
\left.+\frac{r^2 d\theta^2}{P(\theta)}+P(\theta)r^2 \sin^2\theta d\varphi^2\right),
\end{align}
where
\begin{align}
f(r)&=\left(1-\frac{2M}{r}+\frac{Q^2}{r^2}\right)(1-\alpha^2 r^2),\\
P(\theta)&=1-2\alpha M \cos\theta+\alpha^2 Q^2 \cos^2\theta.
\end{align}
The metric is of Petrov type D and has two Killing vectors (the rotational $\partial_\varphi$ and the boost $\partial_t$ Killing vectors). 
The parameters $M$, $Q$ and $\alpha$ are related to the BH mass, charge and acceleration, respectively. The vector potential associated with the electromagnetic field is given by
\begin{equation}
A=-\frac{Q}{r}dt.
\end{equation}
The metric \eqref{Cmetric} asymptotes to the Reissner-Nordstr\"om (RN) solution as $\alpha\rightarrow 0$ and to the $C-$metric as $Q\rightarrow 0$. There is a curvature singularity at $r=0$, while the roots of $f(r)$ determine the causal structure of the spacetime (see Fig. \ref{Penrose}). There exist three null hypersurfaces at 
\begin{eqnarray}
&&r=r_\alpha:=\alpha^{-1},\\
&&r=r_{\pm}:=M\pm\sqrt{M^2-Q^2}, 
\end{eqnarray}
called the acceleration horizon $r_\alpha$, event horizon $r_+$ and Cauchy horizon $r_-$, which must satisfy 
$$r_-\leq r_+\leq r_\alpha.$$ 

We note that the additional feature of accelerating BHs possessing an acceleration horizon, is due to the fact that a uniformly accelerating observer asymptotically approaches the speed of light and, hence, can never observe events beyond that asymptotic light cone. Thus, we limit our attention to the range $r_+<r<r_\alpha$, in which $f(r)$ is positive, and the metric has fixed signature, implying that $P(\theta)>0$ for all $\theta\in\left[0,\pi\right]$. 

For $r_+\leq r_\alpha$ to hold, then 
$$\alpha\leq 1/r_+,$$
where at the equality the BH is extremal (known as the Nariai limit). Additionally, when $M=Q$, the event and Cauchy horizons coincide and the BH is, again, extremal.

Conical singularities generally occur on the axis at $\theta=0$ and $\theta=\pi$, due to the fact that the ratio of the circumference over the radius of the object is not exactly $2\pi$ there, designating the existence of deficit angles. However, by specifying the range of $\varphi$ accordingly, the deficit or excess angle of one of these two conical singularities can be removed. Specifically, if we assume that $\varphi\in  [0,2\pi C )$ and consider the regularity of the half-axis of symmetry $\theta=0$, with $t$ and $r$ constant, then
\begin{equation}
\label{deficit_1}
\frac{\text{circumference}}{\text{radius}}=\lim\limits_{\theta\to 0}\frac{2\pi C P(\theta)\sin\theta}{\theta}=2\pi C P(0),
\end{equation}
where $P(0)=1-2\alpha M +\alpha^2 Q^2$. Equivalently, by considering the regularity at $\theta=\pi$, then
\begin{equation}
\label{deficit_2}
\frac{\text{circumference}}{\text{radius}}=\lim\limits_{\theta\to \pi}\frac{2\pi C P(\theta)\sin\theta}{\pi-\theta}=2\pi C P(\pi),
\end{equation}
where $P(\pi)=1+2\alpha M +\alpha^2 Q^2$. Eqs. \eqref{deficit_1} and \eqref{deficit_2} imply the existence of conical singularities with different conicities. The deficit or excess angles of either of these two conical singularities can be removed by appropriately choosing $C$, but not both simultaneously. In general, $C$ can thus be thought of as determining the balance between the deficit/excess angles on the two parts of the axis.
A natural choice would be to require $C=1/P(\pi)$. The deficit angles at the poles $\theta=0,\,\pi$ are, then,
\begin{equation}
\label{angles_1}
\delta_\pi=0,\,\,\,\,\,\,\,\,\,\,\,\,\,\,\,\delta_0=2\pi\left(1-\frac{P(0)}{P(\pi)}\right),
\end{equation}
which correspond to the removal of the conical singularity at $\theta=\pi$ and the existence of a deficit angle at $\theta=0$. The metric \eqref{Cmetric} can, therefore, be understood as representing a RN-like BH that is being accelerated along the axis $\theta=0$ by the action of a force which corresponds to the tension of a cosmic string \cite{Griffiths:2006tk}. One can, equivalently, remove the conical singularity at $\theta=0$ and obtain an excess angle at $\theta=\pi$ by setting $C=1/P(0)$. In general, there is no need to specify which conical singularity is removed although for the rest of our discussion we will choose 
$$C=\frac{1}{P(\pi)}.$$ 

In the case of the $C-$metric (with $Q=0$), conformal infinity is spacelike for $\theta\in(0,\pi)$ and each black hole has an internal structure which is qualitatively similar to the Schwarzschild BH. The subtleties of the global properties and causal structure of this spacetime are explained in detail in \cite{Griffiths:2006tk}.

In turn, in a maximal analytic extension of the charged $C-$metric, the curvature singularity is timelike and further extensions beyond the Cauchy horizon will, generally, occur. Therefore, the internal structure of such BHs  resemble the RN BH \cite{hawking_ellis_1973}. 
Since the charged $C-$metric is asymptotically flat (in the sense of \cite{ashtekar1981}), a well-defined notion of infinity exists. All observers will intersect the acceleration horizon before reaching infinity, and the causal structure of such solution is, roughly, given by the Penrose diagram in Fig. \ref{Penrose}. The region of the spacetime outside the inner horizon is globally hyperbolic \cite{Hawking:1997ia}.

\begin{figure*}[t]
\subfigure{\includegraphics[scale=0.24]{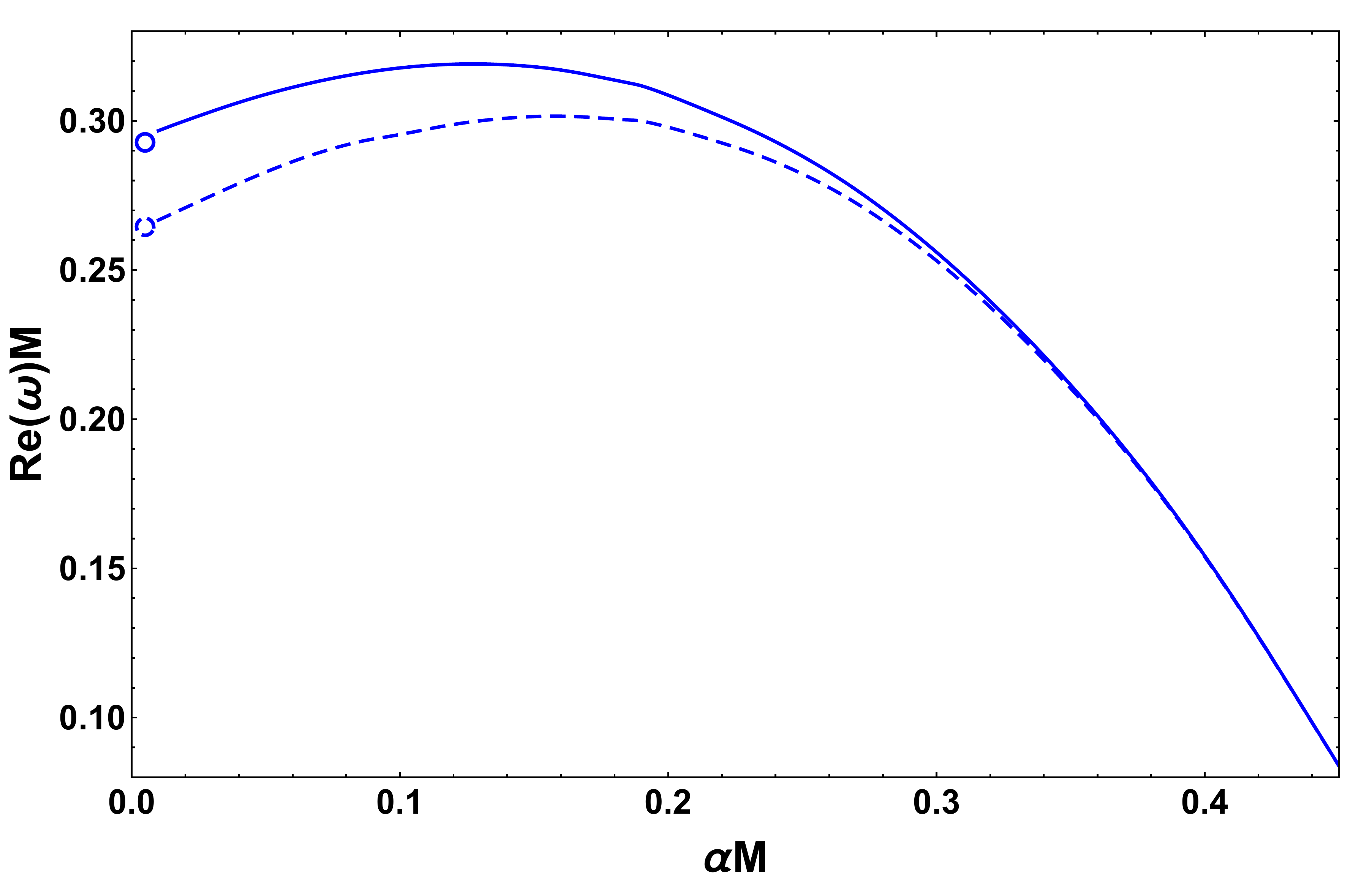}}\hskip 2ex
\subfigure{\includegraphics[scale=0.24]{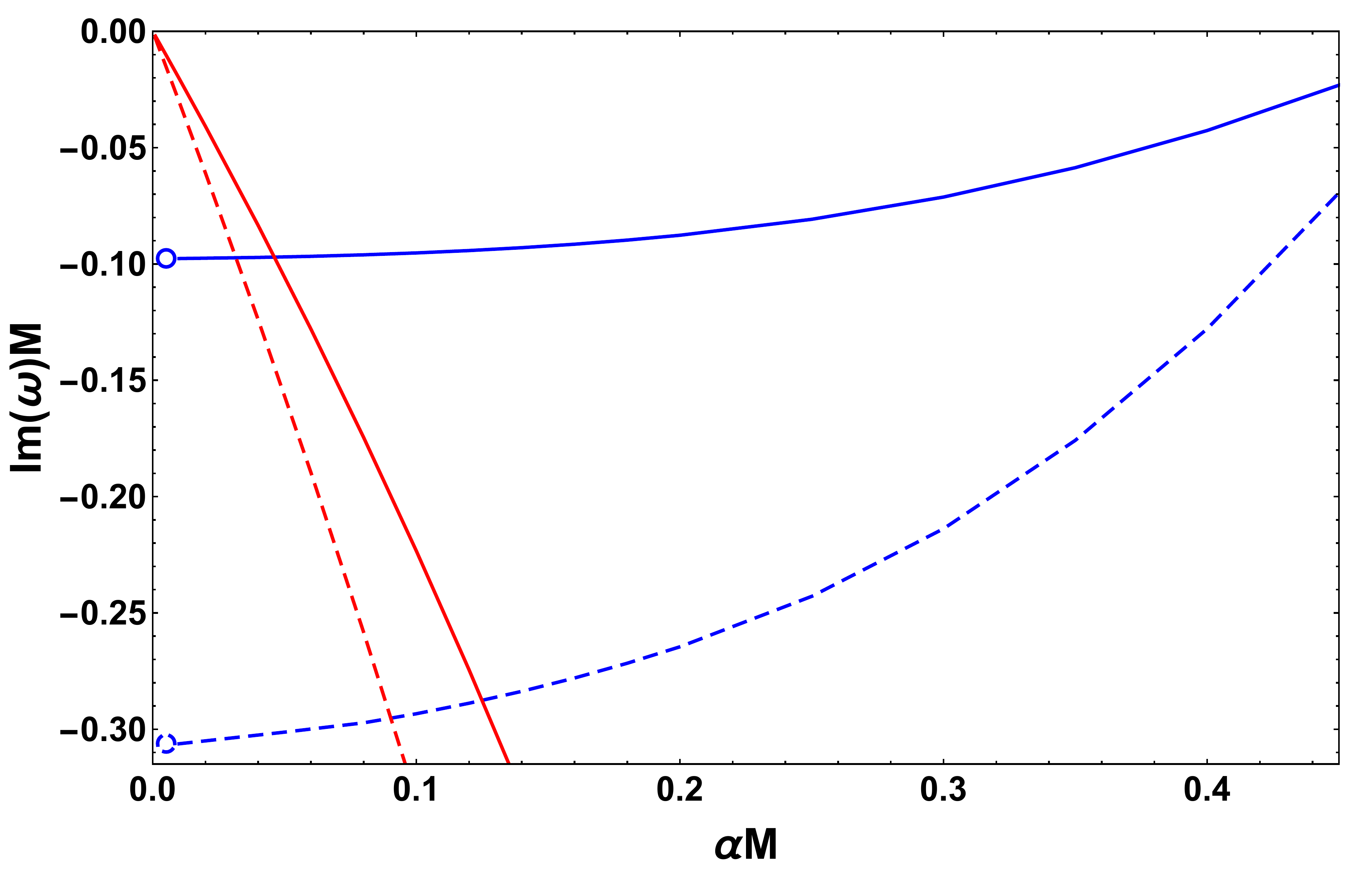}}
\caption{Real (left) and imaginary (right) parts of $n=0$ (solid curves) and $n=1$ (dashed curves) scalar QNMs with $m_0=1$ ($\ell=1$) vs. the boost parameter $\alpha M$ of an accelerating Schwarzschild black hole. The blue curves indicate the photon surface QNMs while the red ones indicate the acceleration modes. The circles at $\alpha=0$ designate the scalar $l=1$ QNMs of the static Schwarzschild black hole.}
\label{Schwarz}
\end{figure*}

Finally, note that we can conformally rescale \eqref{Cmetric} by using the conformal factor 
$$\Omega=1-\alpha r \cos\theta$$ 
and get the conformal metric
\begin{equation}
\label{comp_Cmetric}
d\tilde{s}^2=\Omega^2 ds^2=-f(r)dt^2+\frac{dr^2}{f(r)}+\frac{r^2 d\theta^2}{P(\theta)}+P(\theta)r^2 \sin^2\theta d\varphi^2.
\end{equation}
Such conformal rescaling will prove to be crucial to our study in the following sections.
%%%%%%%%%%%%%%%%%%%%%%%%%%%%%%%%%%%%%%%%
\section{Scalar wave equation}\label{swe}
We consider a massless neutral scalar field $\psi$ minimally coupled to gravity on a spacetime $\mathcal{M}$ with the following matter action 
\begin{equation}
\label{action}
S_m=-\frac{1}{2}\int_\mathcal{M} d^4x\sqrt{-g}\,\partial_\mu\psi\partial^\mu\psi,
\end{equation}
whose dynamical evolution is described by the wave equation
\begin{equation}
\label{wave_eq}
\Box_g \psi=0,
\end{equation}
where we used the standard notation $\Box_g:=g_{\mu\nu}\nabla^{\mu}\nabla^\nu$.

One of the advantages of the $C-$metric is that \eqref{wave_eq} separates (see \cite{Hawking:1997ia}, where this was shown although for a coordinate system different from ours). 

To see this in our coordinate system, we first recall that \eqref{wave_eq} is invariant under the conformal transformation $\tilde{g}_{\mu\nu}\rightarrow \Omega^2 g_{\mu\nu}$, $\tilde{\psi}\rightarrow\Omega^{-1}\psi$ and can be re-written as
\begin{equation}
\label{conf1}
\Box_{\tilde g} \tilde\psi-\frac{1}{6}\tilde{R}\tilde{\psi}=0,
\end{equation}
where $\tilde{R}$ is the conformally rescaled Ricci curvature (recall also that $R=0$ here). For further details of this derivation we refer to \cite{Wald-book}, appendix D.

Then, expanding \eqref{conf1}, by using \eqref{comp_Cmetric}, we find
\begin{align}
\nonumber
&-\frac{r^2\partial^2_t\tilde{\psi}}{f(r)}+\partial_r(r^2f(r)\partial_r\tilde{\psi})
+\frac{1}{\sin\theta}\partial_\theta(P(\theta)\sin\theta\partial_\theta\tilde{\psi})\\\nonumber
&+\frac{\partial^2_\varphi\tilde{\psi}}{\sin^2\theta P(\theta)}+\frac{1}{6}\left(r^2 f^{\prime\prime}(r)+4rf^{\prime}(r)+2f(r)+P^{\prime\prime}(\theta)\right.\\
&\left.+3\cot\theta P^{\prime}(\theta)-2P(\theta)\right)\tilde{\psi}=0,
\label{feq1}
\end{align}
where the prime denotes differentiation with respect to the indicated function variable. 

Remarkably, we can separate \eqref{feq1} by choosing the ansatz 
\begin{equation}
\tilde{\psi}=e^{-i\omega t}e^{i m\varphi}\frac{\phi(r)}{r}\chi(\theta),
\end{equation}
where $\omega$ is the quasinormal frequency and $m$ the magnetic (azimuthal) quantum number. 

Since the azimuthal coordinate $\varphi$ is periodic and the conical singularity along $\theta=\pi$ is removed by requiring $C=1/P(\pi)$, then $m$ must be of the form $m=m_0 P(\pi)$, with $m_0$ integer \cite{Bini:2014kga}. We assume, without loss of generality, that $m_0$ is positive.\footnote{Equivalently, one could rescale the azimuthal coordinate $\varphi=C\Phi$ where $\Phi\in[0,2\pi)$. The line element \eqref{Cmetric}, then, becomes $ds^2=\Omega^{-2}\left(-fdt^2+f^{-1}dr^2+P^{-1}r^2d\theta^2+P C^2 r^2\sin^2\theta d\Phi^2\right)$ and $m$ remains a positive integer without further redefinitions.} 
\begin{figure*}[t]
\subfigure{\includegraphics[scale=0.24]{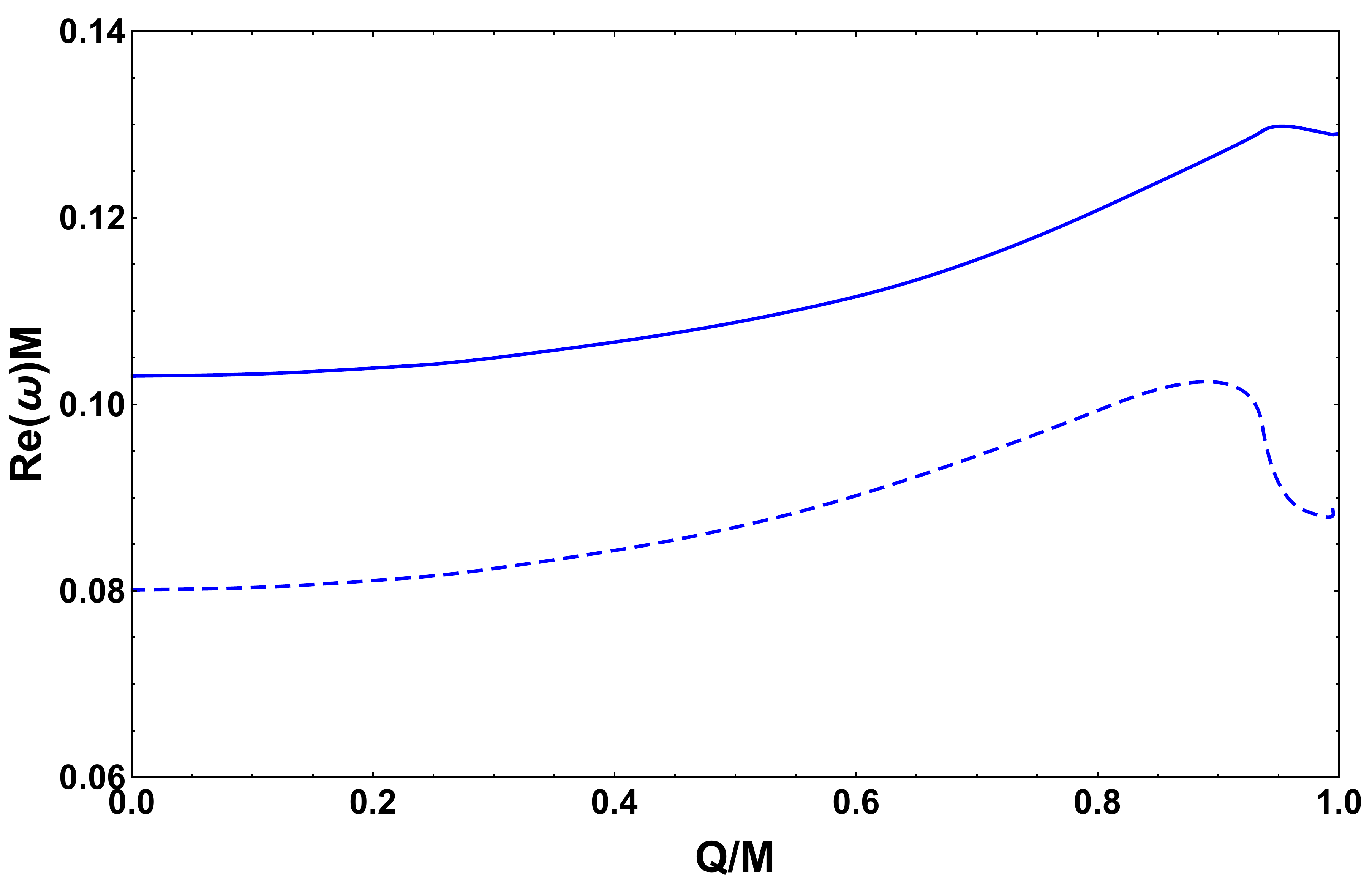}}\hskip 2ex
\subfigure{\includegraphics[scale=0.24]{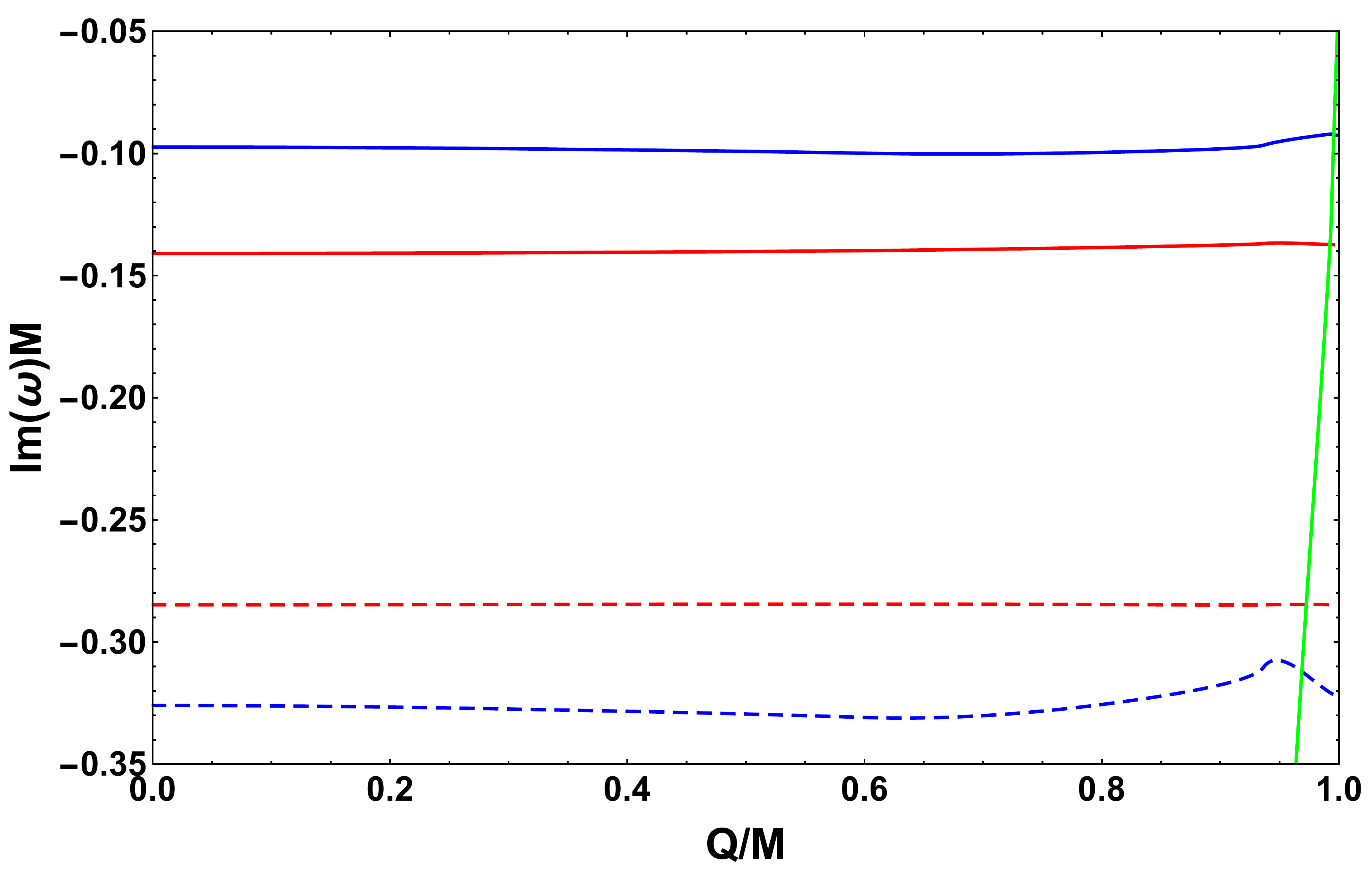}}
\caption{Real (left) and imaginary (right) parts of $n=0$ (solid curves) and $n=1$ (dashed curves) scalar QNMs with $m_0=0$ ($\ell=0$) vs. the electric charge $Q/M$ of an accelerating Reissner-Nordstr\"om black hole with boost parameter $\alpha M=0.13$. The blue curves indicate the photon surface QNMs, the red ones indicate the acceleration modes, while the green indicate the near extremal modes.}
\label{RNQ}
\end{figure*}
Consequently, \eqref{conf1} becomes
\begin{eqnarray}
\label{final_radial}
\frac{d^2\phi(r)}{dr^2_*}+(\omega^2-V_r)\phi(r)&=0,\\
\label{final_polar}
\frac{d^2\chi(\theta)}{dz^2}-(m^2-V_\theta)\chi(\theta)&=0,
\end{eqnarray}
where
\begin{equation}
dr_*=\frac{dr}{f(r)},\,\,\,\,\,\,\,\,\,\,\,\,\,dz=\frac{d\theta}{P(\theta)\sin\theta},
\end{equation}
and
\begin{align}
\label{pot_r}
V_r&=f(r)\left(\frac{\lambda}{r^2}-\frac{f(r)}{3r^2}+\frac{f^\prime(r)}{3r}-\frac{f^{\prime\prime}(r)}{6}\right),\\\nonumber
V_\theta&=P(\theta)\left(\lambda \sin^2\theta-\frac{P(\theta)\sin^2\theta}{3}+\frac{\sin\theta\cos\theta P^\prime(\theta)}{2}\right.\\
\label{pot_theta}&\,\,\,\,\,\,\,\,\,\,\,\,\,\,\,\,\,\,\,\,\,\,+\left.\frac{\sin^2\theta P^{\prime\prime}(\theta)}{6}\right).
\end{align}
We note that equation (\ref{final_radial}) is equivalent to equation (13) in \cite{Bini:2014kga}, taking into account the proper limits, $a \rightarrow 0$, $Q \rightarrow 0$ and $-2K \rightarrow \lambda -1/3$ which, in the Schwarzschild limit, matches the corresponding $K$ in \cite{Bini:2014kga}.

In order to solve the eigenvalue problems relative to \eqref{final_radial} and \eqref{final_polar}, we need to apply appropriate boundary conditions. For QNMs, the boundary conditions are divided in two categories for $\phi(r)$ and $\chi(\theta)$. The physically motivated boundary conditions in this case are 
\begin{align}
\label{bcs_radial}
\phi(r) &\sim
\left\{
\begin{array}{lcl}
e^{-i\omega r_* },\,\,\,\quad r_*\rightarrow-\infty \,\,(r \rightarrow r_+),\\
&
&
\\
 e^{+i\omega r_*},\,\,\,\quad r_*\rightarrow+\infty\,\,(r \rightarrow r_\alpha),
\end{array}
\right.\\
\label{bcs_polar}
\chi(\theta) &\sim
\left\{
\begin{array}{lcl}
e^{+mz},\,\,\,\quad z\rightarrow-\infty \,\,(\theta \rightarrow 0), \\
&
&
\\
 e^{-mz},\,\,\,\quad z\rightarrow+\infty\,\,(\theta \rightarrow \pi).
\end{array}
\right.
\end{align}\\
Physically, conditions \eqref{bcs_radial} mean that beyond the event horizon, all events are causally disconnected with the external region (purely ingoing waves), while beyond the acceleration horizon incoming waves are unobservable (purely outgoing waves). In turn, conditions \eqref{bcs_polar} are obviously taken so that our solution is physically interesting and the scalar field does not blow up at the interval boundaries of $\theta$. 

The striking advantage of the charged $C-$metric is the absence of coupled $\mathcal{O}(\omega\lambda)$ terms in \eqref{final_radial} and \eqref{final_polar}. This suggests that one can solve \eqref{final_polar} with \eqref{bcs_polar} to obtain the eigenvalues $\lambda$ associated to some given angular and magnetic quantum numbers $\ell$ and $m$, respectively. 

Due to the conformal coupling, at the limit $\alpha\rightarrow 0$ the separation constant takes the exact form $\lambda=\ell(\ell+1)+1/3$ \cite{Kofron_2015}. Hence, we can map each $\lambda$ obtained by solving \eqref{final_polar} to a certain $\ell$.\footnote{E.g., in the limit $\alpha\rightarrow 0$, the value $\lambda=1/3$ corresponds to $\ell=0$, while $\lambda=7/3$ corresponds to $\ell=1$ and so on.} 

Finally, by solving \eqref{final_radial} with \eqref{bcs_radial} together with a separation constant calculated previously, we obtain a discrete set of QNMs $\omega$ depending on the choice of the BH parameters, magnetic quantum number $m$, separation constant $\lambda$ and overtone number $n$, where the $n=0$ mode is called the fundamental/dominant QNM. The fundamental QNM, therefore, is the one that has the smallest (in absolute value) imaginary part which will decay last and will dominate the late-time behavior of the ringdown waveform.

In the following sections, the procedure that we will use to integrate equations \eqref{final_radial} and \eqref{final_polar} begins by the calculation of the eigenvalues $\lambda$, given specific $M,Q,\alpha$ and $m_0$. This will be achieved by using the methods described below to compute QNMs and checked with a Frobenius-like method, whose main steps are revised in Appendix \ref{appb}.

To calculate the QNMs, we will use the \textit{Mathematica} package \textit{QNMSpectral} developed in \cite{Jansen:2017oag}, which is based on the discretization of the differential equations using pseudospectral collocation methods and directly solving the resulting generalized eigenvalue problem. We also perform time-domain evolutions of the equations by using the method developed in \cite{Gundlach:1993tp}, where the wave equation is integrated in double-null coordinates using two Gaussian wave-packets as initial data. We extract the QNMs by applying the Prony method on the numerically evolved perturbation \cite{Berti:2007dg}. We justify our findings by a direct comparison of the resulting QNMs from the aforementioned methods (see Table \ref{table_comp} in Appendix \ref{appa}) and by further cross-checking the results at the eikonal limit with the Wentzel-Kramers-Brillouin (WKB) method \cite{WKB} as well as by calculating the instability timescale of null geodesics at the photon surface.
\begin{figure*}[t]
\subfigure{\includegraphics[scale=0.24]{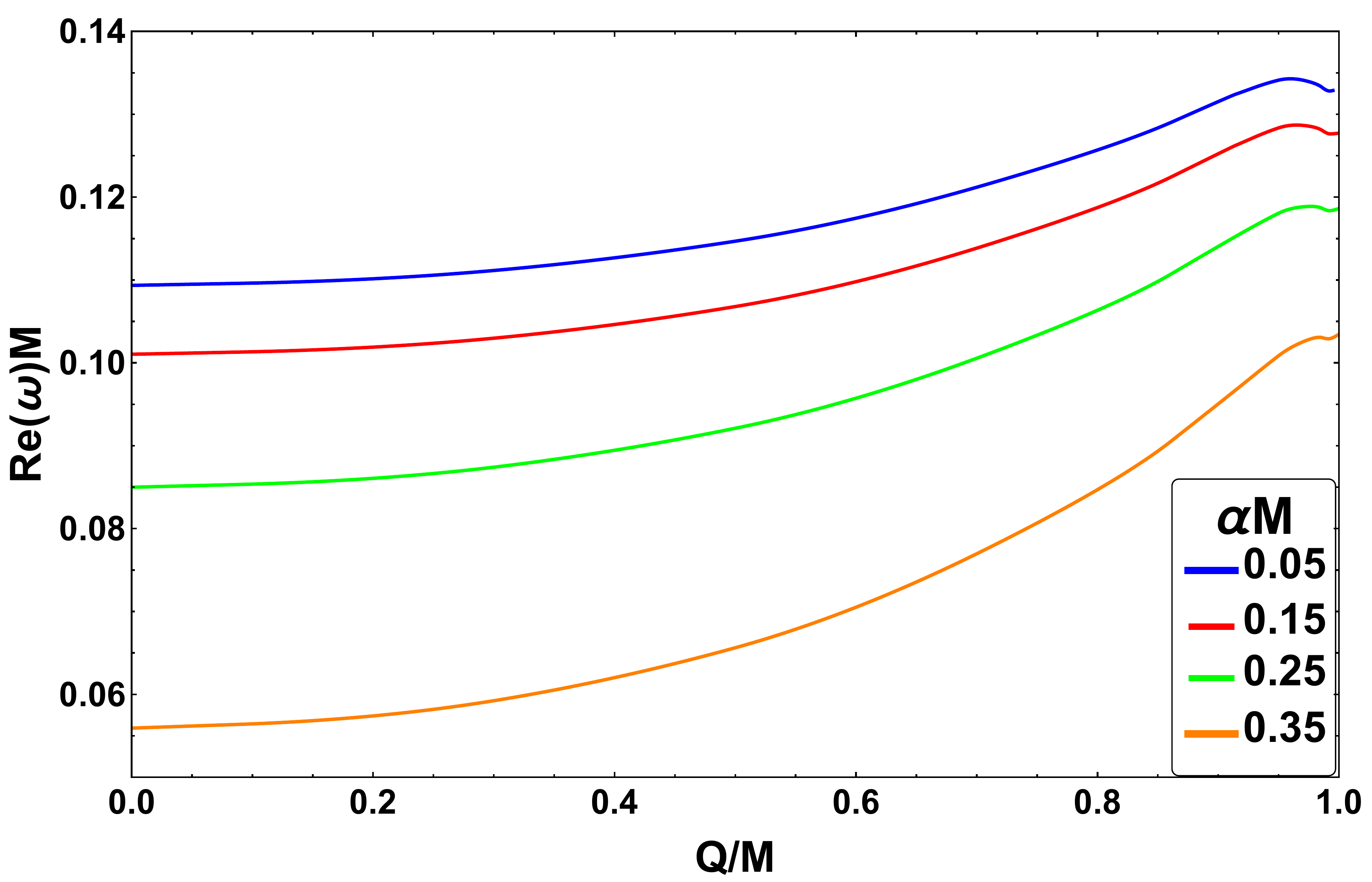}}\hskip 2ex
\subfigure{\includegraphics[scale=0.24]{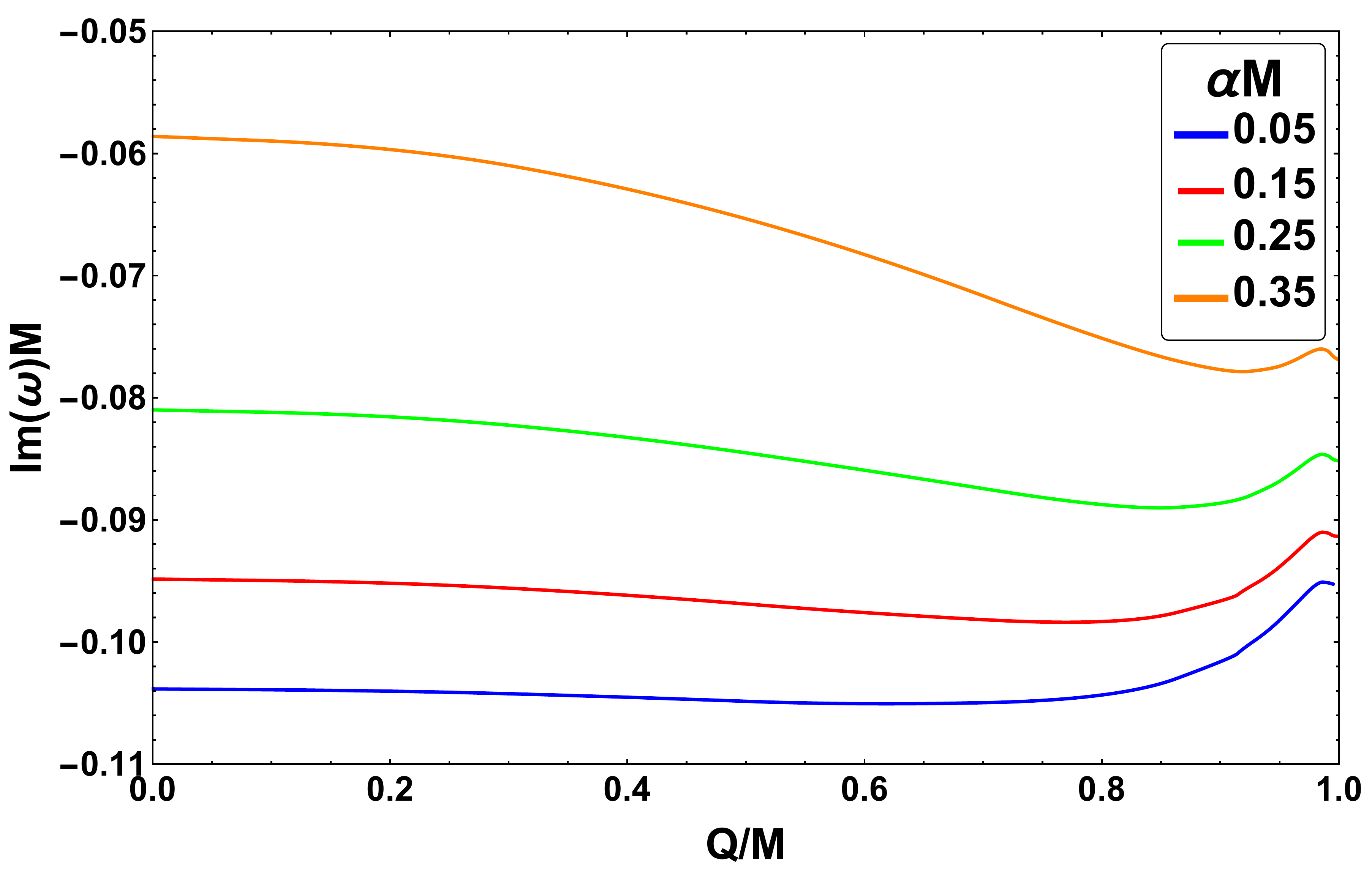}}
\subfigure{\includegraphics[scale=0.24]{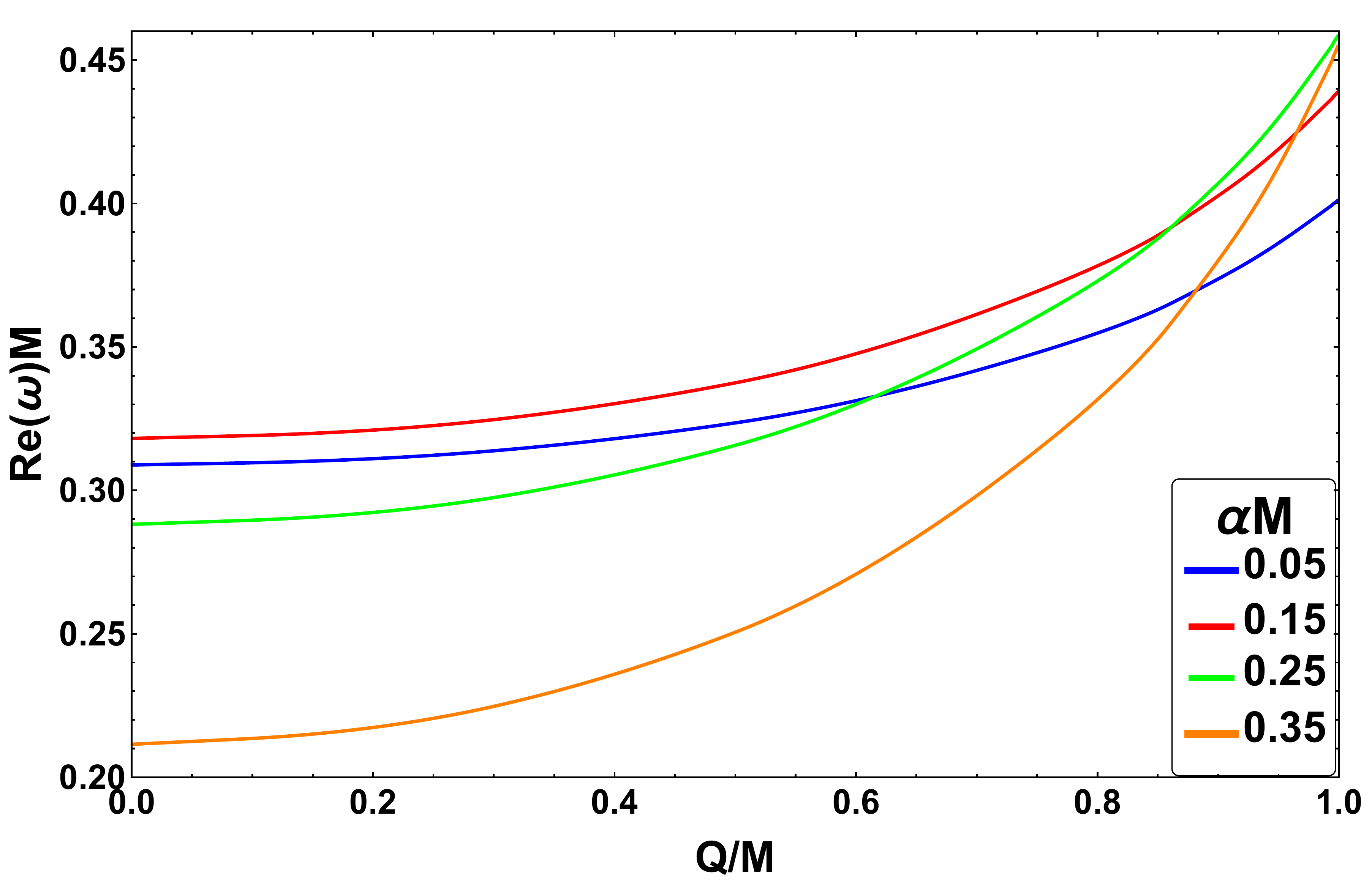}}\hskip 2ex
\subfigure{\includegraphics[scale=0.24]{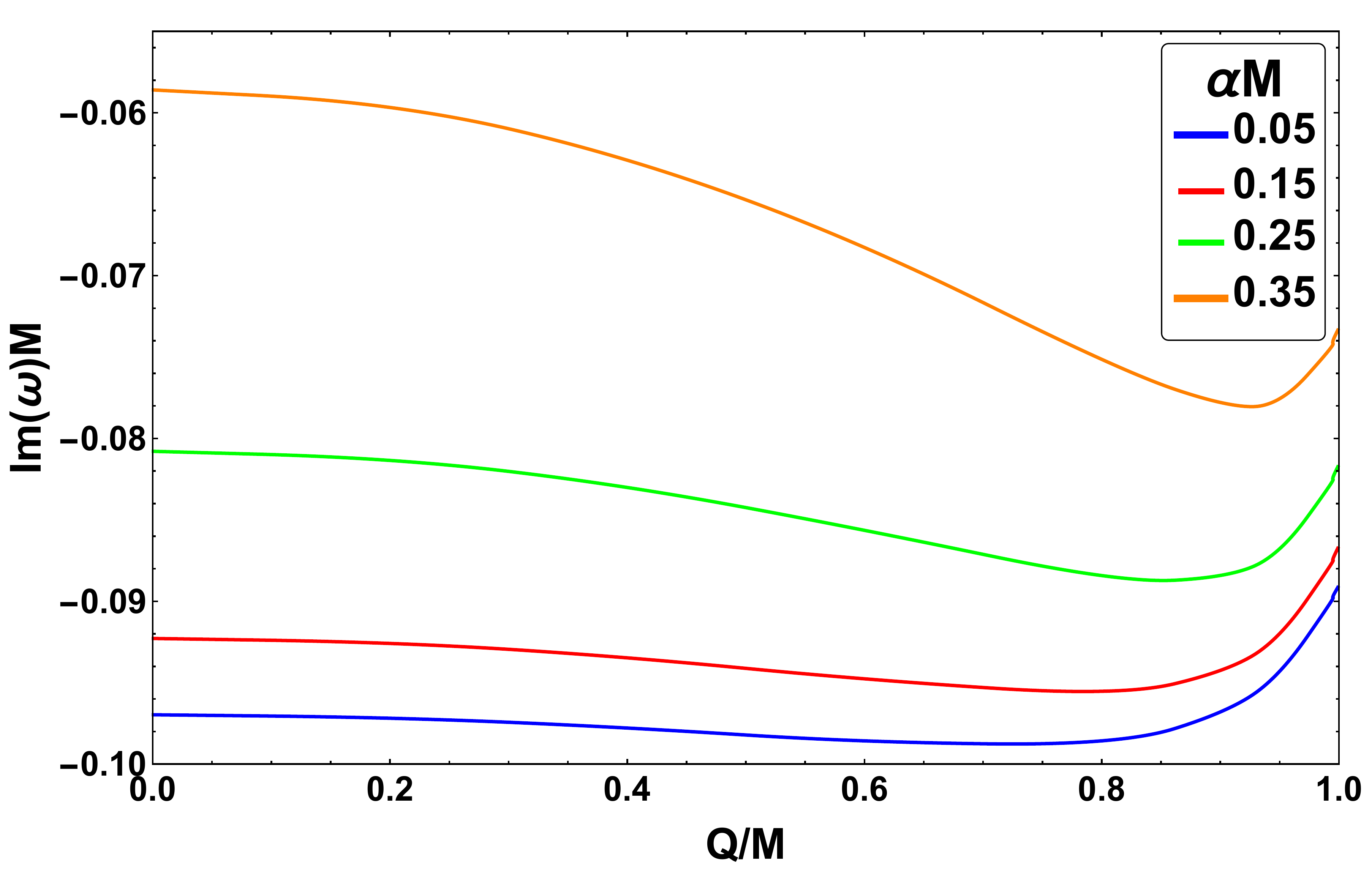}}
\caption{Real (left) and imaginary (right) parts of $n=0$ photon surface QNMs with $m_0=0$ ($\ell=0$) (top panel) and $m_0=1$ ($\ell=1$) (bottom panel) vs. the electric charge $Q/M$ of an accelerating Reissner-Nordstr\"om black hole, for various boost parameters $\alpha M$.}
\label{Qa}
\end{figure*}
%%%%%%%%%%%%%%%%%%%%%%%%%%%%%%%%%%%%%%%%%%%%%%%%%%
\section{Quasinormal modes of accelerating black holes: The distinct families}
Our numerics indicate that there exist three distinct families of scalar QNMs in the charged $C-$metric. These modes antagonize each other, meaning that for different regions in the BH parameter space, different families dominate the ringdown signal against the other. We now discuss their behavior and dependence on the various parameters which define our physical system.
\\\\
\noindent{\bf Photon surface (PS) modes $\mathbb{\omega}_\text{PS}$:}
Many BH spacetimes possess a photon sphere, i.e. a region where null particles are trapped in unstable circular orbits.\footnote{Theorems about the existence and uniqueness of the photon sphere in static asymptotically flat spacetimes can be found in \cite{Cederbaum1, Cederbaum2}. However, those results do not include boosted black holes solutions.}. 
It has been shown in \cite{Cardoso:2008bp} that the photon sphere, and its geometrical properties, have a strong pull on the decay of perturbations in BH solutions of GR (for other theories, see \cite{Konoplya:2017wot}). In fact, the orbital frequency and instability timescale of null geodesics at the photon sphere are directly associated with the real and imaginary parts, respectively, of QNMs at the eikonal limit, where $m\sim \ell\gg 1$. 

Since in our case there is an aspherical photon surface\footnote{See \cite{Gibbons-Warnick} for a study of the photon surface in the $C-$metric.}, we call these modes ``photon surface" QNMs. We distinguish the PS modes from the rest due to their oscillatory nature. The $C-$metric possesses a photon surface with a ``radius" that, at the equatorial plane, coincides with the peak of the effective potential \eqref{pot_r} for $\lambda\sim m \sim\ell\to\infty$. 

\begin{figure*}[t]
\subfigure{\includegraphics[scale=0.24]{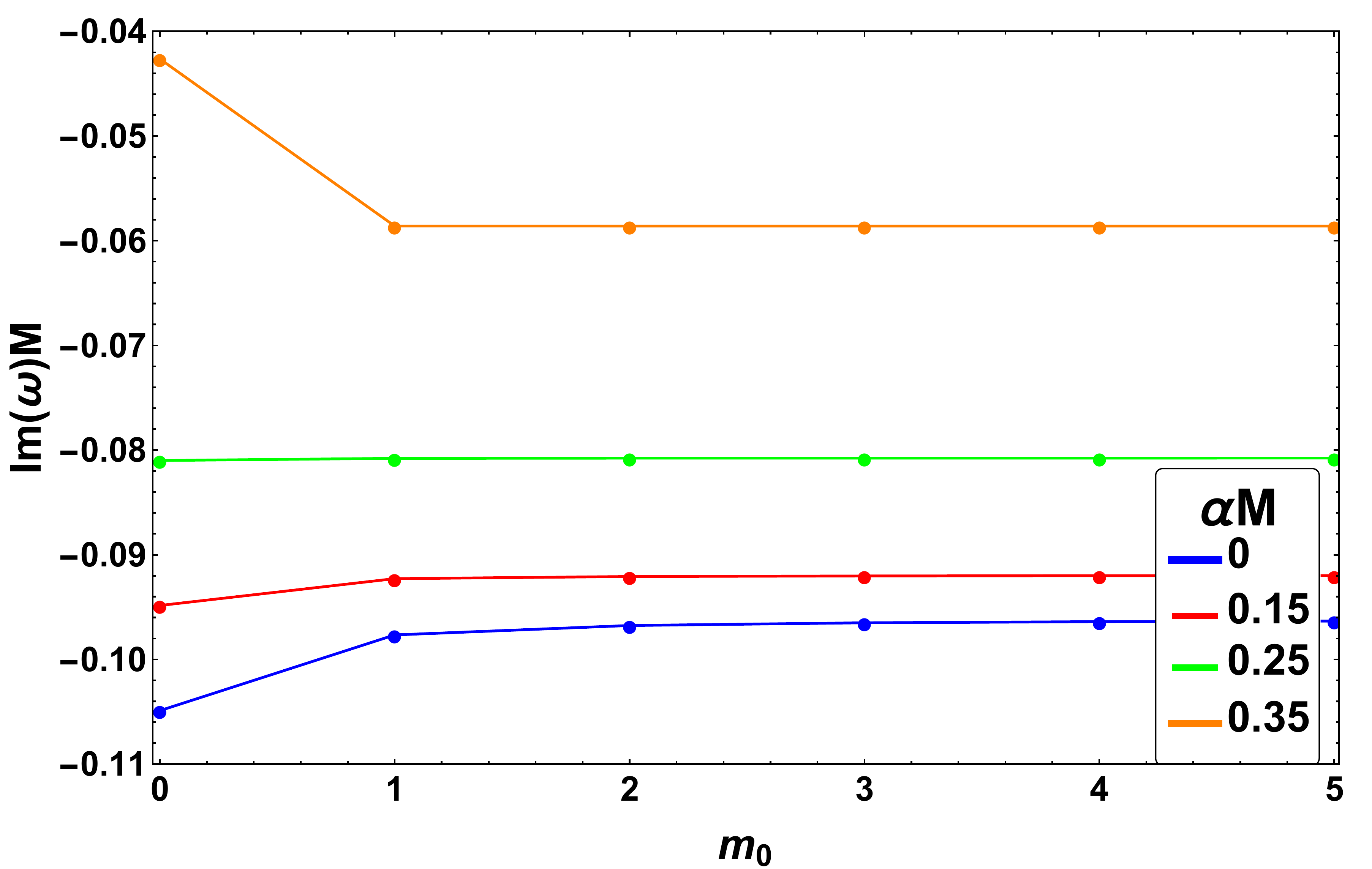}}\hskip 2ex
\subfigure{\includegraphics[scale=0.24]{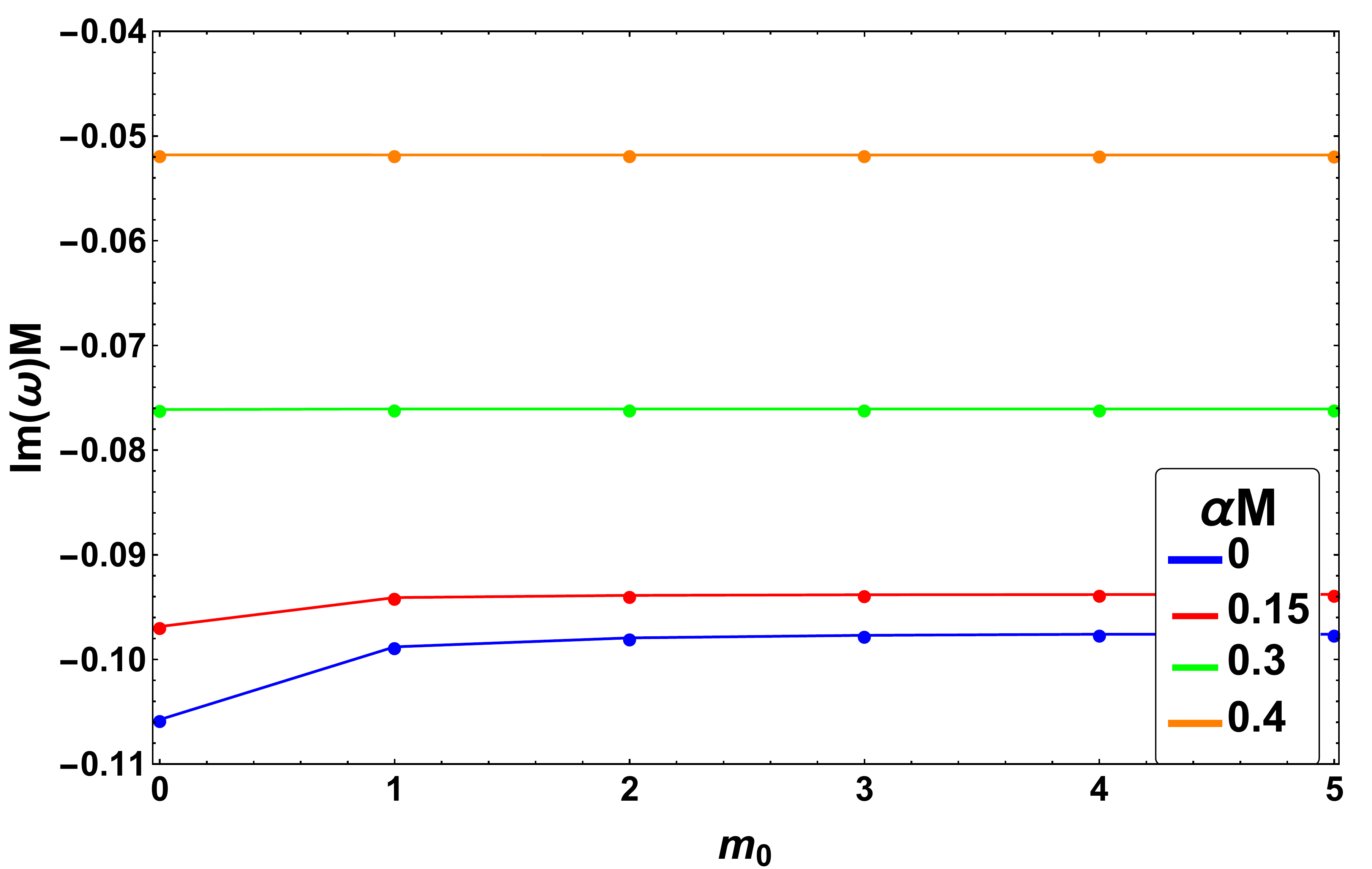}}
\caption{Imaginary part of the fundamental photon surface modes of the $C-$metric (left) and the charged $C-$metric with $Q/M=0.5$ (right) as a function of $m_0$ and $\alpha M$. Recall that when $\alpha=0$, then $m_0=\ell$.}
\label{anomalous}
\end{figure*}

\begin{figure}[h]
\subfigure{\includegraphics[scale=0.23]{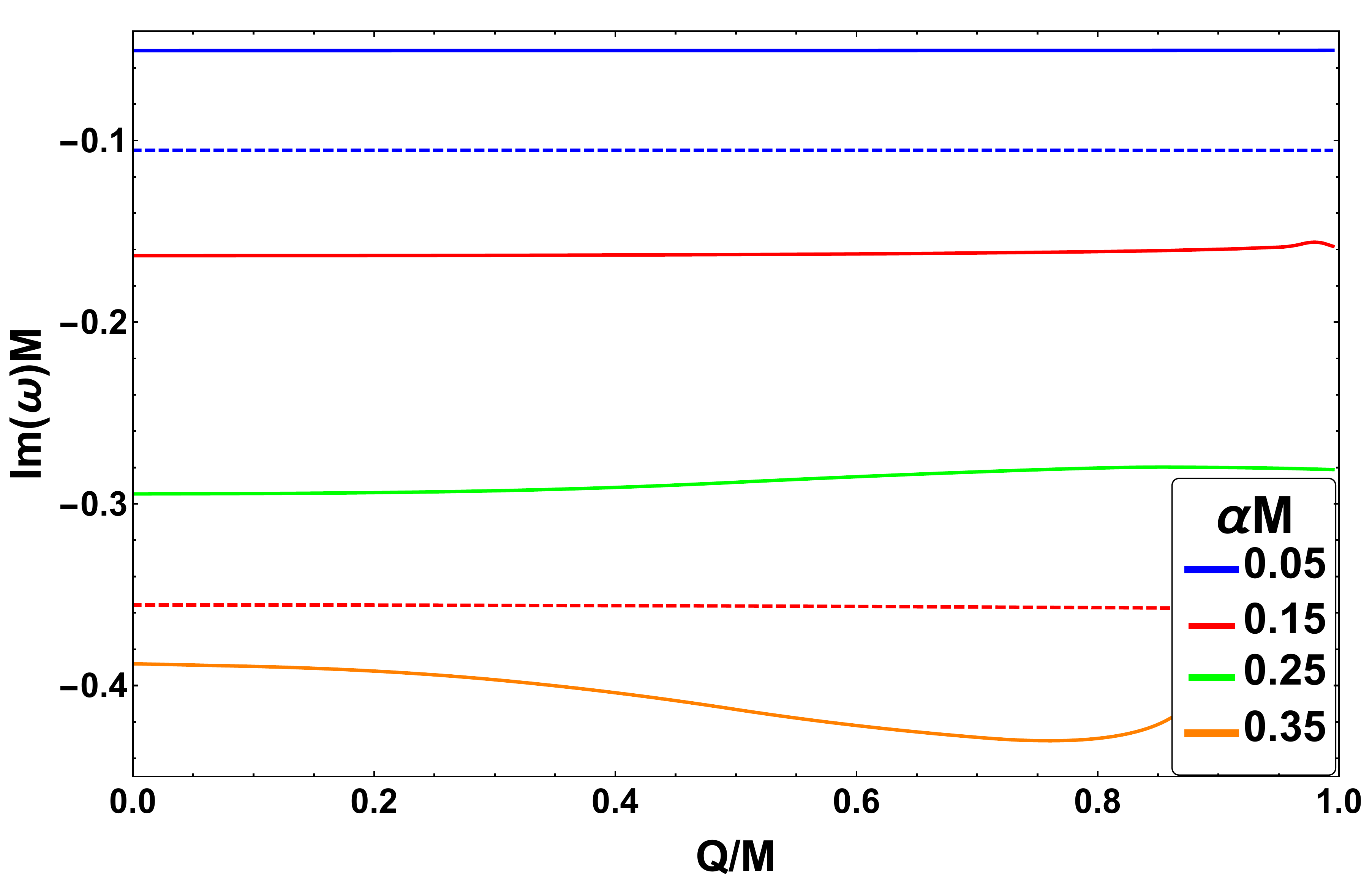}}
\caption{Imaginary parts of $n=0$ acceleration modes with $m_0=0$ ($\ell=0$) (solid lines) and $m_0=1$ (dashed lines) vs. the electric charge $Q/M$ of an accelerating Reissner-Nordstr\"om black hole for various boost parameters $\alpha M$.}
\label{acc}
\end{figure}

In Fig. \ref{Schwarz}, we show the effect of the parameter $\alpha M$ on the real and imaginary parts of the PS modes for an accelerating Schwarzschild BH. The real part grows until it reaches a maximum as $\alpha M$ increases, and then decreases. The imaginary part decreases (in absolute value) as the boost increases, indicating that the modes will become more dominant at the ringdown waveform. As the boost parameter tends to its extremal value $\alpha=1/r_+$ (usually referred to as the Nariai limit \cite{Dias:2003up}) both real and imaginary parts approach $0$. This result could be anticipated if one considers the fact that the Nariai modes of de Sitter BHs are proportional to the surface gravity of the event (or cosmological) horizon \cite{Cardoso:2003sw,Molina:2003ff}, which vanishes at the limit $\Lambda\rightarrow \Lambda_\text{max}$, where $\Lambda_\text{max}$ is the extremal cosmological constant, and that both the cosmological and acceleration horizons have quite similar geometric properties. 

We note that at the limit where the acceleration vanishes, the PS QNMs (together with the corresponding eigenvalues and eigenfunctions) asymptote to the Schwarzschild BH scalar QNMs, which are designated with circles in Fig. \ref{Schwarz}.

In Fig. \ref{RNQ} and \ref{Qa}, we turn our attention to the effect of the acceleration on the real and imaginary parts of the PS modes for an accelerating RN BH. For $m_0=0$, it is evident that the real part tends to increase with $Q/M$, for most of the parameter space (there seems to be a turning point close to extremality, but there is no significant decrement beyond that point). The increment on the acceleration, on the other hand, tends to decrease the frequencies (real part) of the QNMs. The absolute value of the imaginary part is affected in the same way for all $m_0$. For $m_0>0$, the real part is a monotonic increasing function of $Q/M$, while as $\alpha M$ grows then a behavior similar to Fig. \ref{Schwarz} is observed. 

\begin{table}[h]
\centering
\scalebox{0.8}{
\begin{tabular}{||c| c | c ||} 
\hline
  \multicolumn{3}{||c||}{$Q/M=0.3$} \\
   \hline
    $m_0 $ & $\alpha M=0.05$ & $\alpha M=0.1$  \\ [0.5ex] 
  \hline
    &$\omega_\alpha$=-0.0506 i  &$\omega_\text{PS}$=0.1079 - 0.1012 i   \\
  0 & $\omega_\text{PS}$=0.1112 - 0.1042 i &$\omega_\alpha$=-0.1055 i\\ 
   & $\lambda$=0.3317&$\lambda$=0.3269\\ 
   \hline
     &$\omega_\text{PS}$=0.3139 - 0.0974 i  &$\omega_\text{PS}$=0.3234 - 0.0957 i   \\
    1  & $\omega_\alpha$=-0.1054 i &$\omega_\alpha$=-0.2234 i\\ 
      & $\lambda$=2.6461&$\lambda$=2.9847\\ 
   \hline
    &$\omega_\text{PS}$=0.5246 - 0.0967 i  &$\omega_\text{PS}$=0.5454 - 0.0953 i  \\
    2  &$\omega_\alpha$=-0.1605 i  &$\omega_\alpha$=-0.3433 i\\  
      & $\lambda$=7.3755&$\lambda$=8.4983\\
   \hline
    & $\omega_\text{PS}$=2.2223 - 0.0963 i &$\omega_\text{PS}$=4.5673 - 0.0951 i\\ 
    10  & -$\lambda_0/2$=-0.0963  &-$\lambda_0/2$=-0.0951 \\ 
       & $\lambda$=132.1497&$\lambda$=155.4315\\
   \hline
\end{tabular}}
\scalebox{0.8}{
\begin{tabular}{||c| c | c ||} 
\hline
  \multicolumn{3}{||c||}{$Q/M=0.999$} \\
   \hline
    $m_0 $  & $\alpha M=0.3$ & $\alpha M=0.5$ \\ [0.5ex] 
  \hline
   &$\omega_\text{NE}$=-0.0412 i  & $\omega_\text{NE}$=-0.0342 i \\ 
  0 & $\omega_\text{PS}$=0.1117 - 0.0814 i &$\omega_\text{PS}$=0.0731 - 0.0581 i \\
   &$\lambda$=0.3033  & $\lambda$=0.2497\\ 
   \hline
       &$\omega_\text{PS}$=0.4596 - 0.0780 i & $\omega_\text{PS}$=0.4016 - 0.0549 i\\
    1    &$\omega_\text{NE}$=-0.1147 i & $\omega_\text{NE}$=-0.1232 i\\ 
     &$\lambda$=4.8970  & $\lambda$=7.4893\\ 
   \hline
    & $\omega_\text{PS}$=1.8128 - 0.0778 i &$\omega_\text{PS}$=1.6043 - 0.0548 i\\ 
    5  & -$\lambda_0/2$=-0.0778  &-$\lambda_0/2$=-0.0548 \\ 
       & $\lambda$=75.8762 &$\lambda$=119.4207\\
   \hline
\end{tabular}}
\caption{Scalar QNMs of the charged $C-$metric for various parameters, where $\lambda$ are the separation constants, $\lambda_0$ are the instability timescales of null geodesics, $\omega_\text{PS}$ the photon surface QNMs, $\omega_\alpha$ the acceleration modes and $\omega_\text{NE}$ the near extremal modes. The absence of acceleration modes for large $m_0$ and $\alpha M$ is due to the fact that they are too subdominant to be captured by our numerics. The same reasoning applies to the absence of near extremal modes for small $Q/M$ and large $m_0$.}
\label{table}
\end{table}

The longest-lived massless QNMs of non-accelerating BHs are those with high angular number $\ell$ and this still holds true when the scalar field is light. On the other hand, if the scalar field is heavy, the longest-lived QNMs are those with low angular number $\ell$ \cite{Lagos:2020oek,Aragon:2020tvq}. 

For an accelerating BH, we observe a similar phenomenon where the defining factor of the ``anomaly" is not the mass but rather the acceleration. In Fig. \ref{anomalous}, we show the anomalous behavior of PS modes. For small $\alpha M$, the dominant modes are obtained for large $m\sim\ell$ (see also Tables \ref{table} and \ref{table2}), while for large $\alpha M$, the dominant modes are obtained for $m_0=\ell=0$. 

The addition of electric charge saturates significantly this anomaly (see the left plot of Fig. \ref{anomalous}), which still occurs but now the threshold $\alpha M$ is significantly higher compared to the $Q=0$ case. Our investigation shows that as the BH charge becomes extremal, this effect becomes even more suppressed.

In Tables \ref{table} and \ref{table2}, we have validated our results by comparing the modes from our numerics with the instability timescale of null geodesics at the equatorial plane of the photon surface, where one can realize that convergence is achieved sufficiently fast as one increases $m$. It is apparent that the PS family will dominate against the rest of the modes, for most of the parameter space and for sufficiently large acceleration parameters, as discussed below.
\begin{figure*}[t]
\subfigure{\includegraphics[scale=0.24]{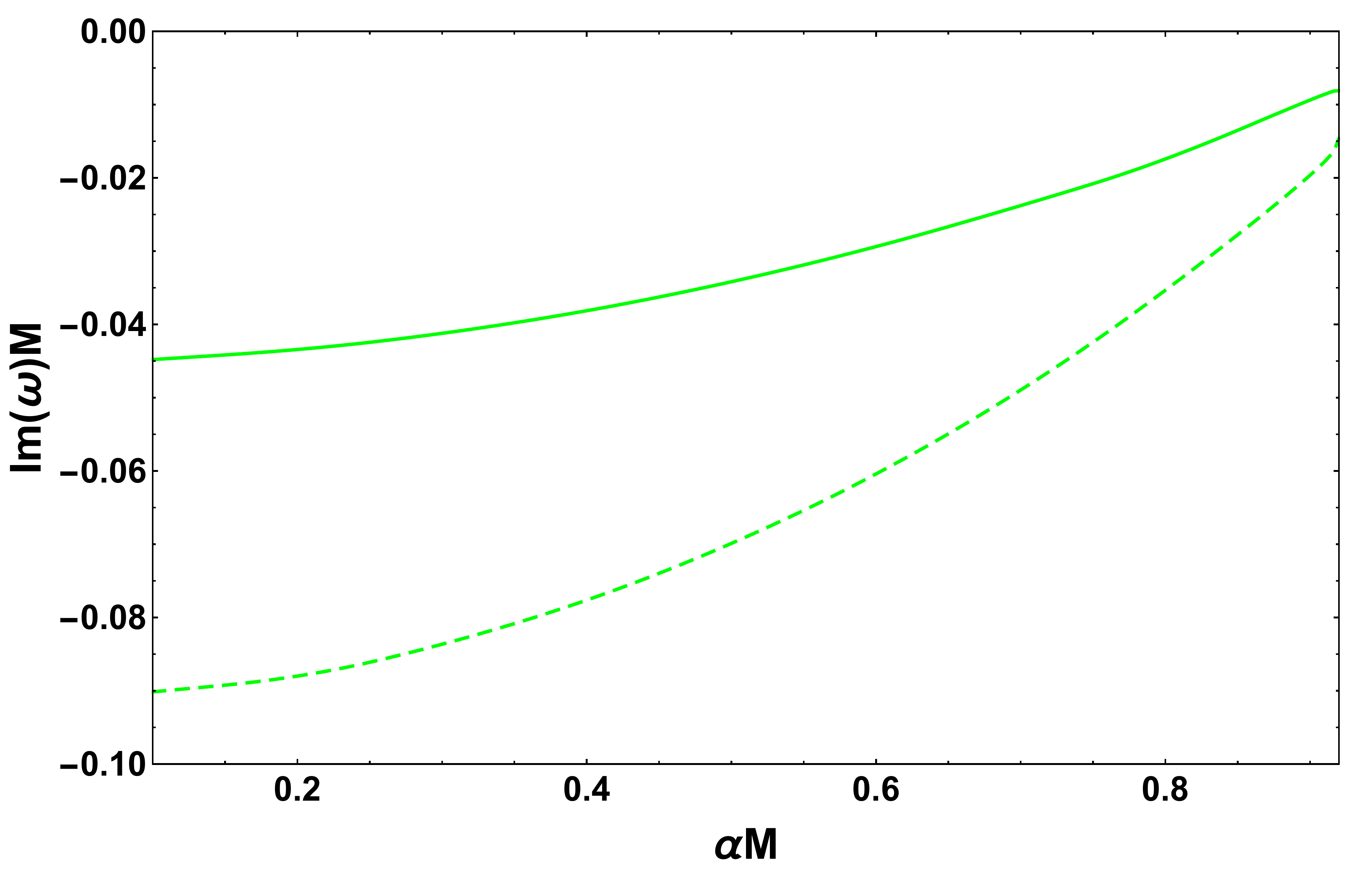}}\hskip 2ex
\subfigure{\includegraphics[scale=0.24]{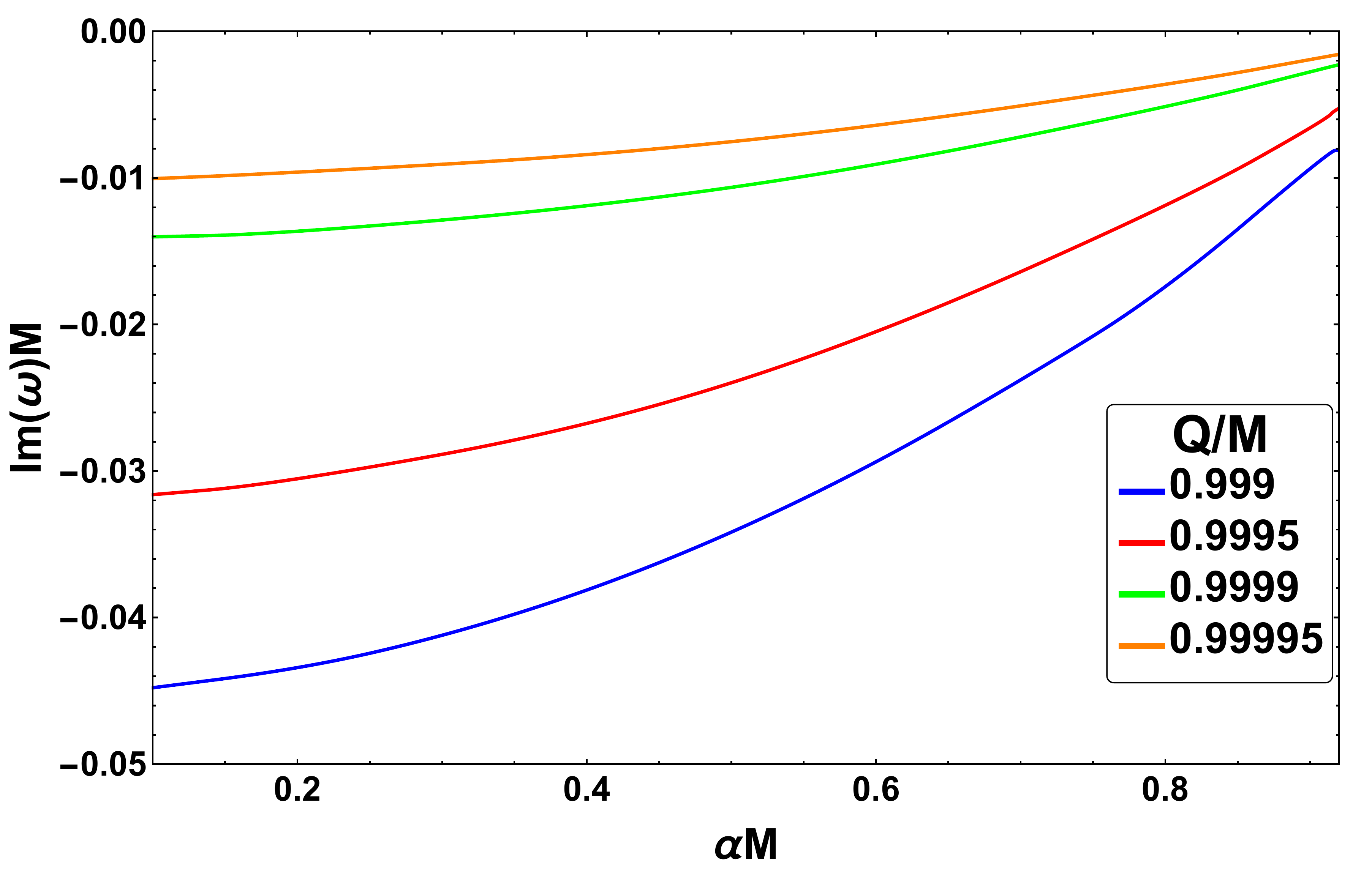}}
\caption{Left: $n=0$ (solid curves) and $n=1$ (dashed curves) near-extremal modes with $m_0=0$ ($\ell=0$) vs. the boost parameter $\alpha M$ of an accelerating Reissner-Nordstr\"om black hole with electric charge $Q/M=0.999$. Right: Fundamental near-extremal modes with $m_0=0$ ($\ell=0$) vs. the boost parameter $\alpha M$ of an accelerating Reissner-Nordstr\"om black hole with varying electric charge $Q/M$.}
\label{Qmax}
\end{figure*}
\\\\
\noindent{\bf Acceleration modes $\mathbb {\omega_\alpha}$:} 
This novel family consists of purely imaginary modes which grow rapidly (in absolute value) as the boost parameter increases (see Fig. \ref{Schwarz} and Fig. \ref{acc}).  
Such modes are absent in non-accelerating spacetimes, so their existence depends solely on the acceleration horizon of the boosted spacetime. 

The acceleration modes seem to follow a linear dependence on the surface gravity of Rindler space $\kappa^\text{R}_\alpha=\alpha$. For sufficiently small boosts 
\begin{equation}
\label{acc_modes}
\omega_\alpha\simeq-\kappa^\text{R}_\alpha(m+n+k+1)i,
\end{equation}
where $k=\ell-m_0$. Higher overtones exhibit larger deformations with respect to \eqref{acc_modes}. It is very interesting that these modes have a weak dependence on the BH charge and only depend on the surface gravity of Rindler space instead of the surface gravity of the BH in study. 

The acceleration modes found here share many similarities with the de Sitter modes found in \cite{Jansen:2017oag,Cardoso:2017soq} and further analyzed in \cite{Cardoso:2018nvb,Destounis:2018qnb,Liu:2019lon,Destounis:2019hca,Destounis:2019omd}. As in \eqref{acc_modes}, the de Sitter modes are also purely imaginary, depend on the surface gravity of pure de Sitter space and have a weak dependence on the BH charge.

In Fig. \ref{acc} we demonstrate how weakly this family of modes depends on $Q/M$. The absolute value of the imaginary part grows rapidly with $\alpha M$, making the modes difficult to be captured numerically for large $\alpha M$. Nevertheless, we can clearly distinguish how strong the effect of the boost is, especially for modes with $m_0>0$.

It is evident that the acceleration modes will become the dominant family for $m=\ell=0$ and small boost parameters (see Table \ref{table} and \ref{table2}) and will play an important role at the late-time behavior of the ringdown waveform, in that parameter region. 
\\\\
\noindent{\bf Near extremal modes $\mathbb{\omega_\text{NE}}$:} Besides the modes already discussed, a third family of modes appear in Fig. \ref{RNQ} (with green color) when the BH is charged. We call this family ``near extremal" (NE) since it only arises when the Cauchy and event horizon approach each other. 
Such modes have been found, analytically \cite{Hod:2017gvn} and numerically \cite{Cardoso:2017soq,Cardoso:2018nvb,Destounis:2018qnb,Liu:2019lon,Destounis:2019hca,Destounis:2019omd}, in the context of non-accelerating charged black holes and play an important role on the dynamics of the ringdown waveform when $Q\rightarrow Q_\text{max}$, where $Q_\text{max}$ is the extremal electric charge of the BH. They arise from $-\infty$ (see Fig. \ref{RNQ}) and at extremality they become zero modes. 

At the limit $r_-\rightarrow r_+$, we can approximate these modes by the following equation
\begin{equation}
\omega_\text{NE}\simeq-\kappa_-(m+n+k+1) i=-\kappa_+(m+n+k+1) i,
\end{equation}
where $k=\ell-m_0$ and $\kappa_\pm=|f^\prime(r_\pm)|/2$ is the surface gravity \cite{Griffiths:2006tk,Gregory:2017ogk} of either the event horizon $r_+$ or the Cauchy horizon $r_-$. 

In Fig. \ref{Qmax} we observe that as $\alpha M$ increases, the NE modes decay slower, while as we approach extremality, the absolute value of their imaginary parts becomes even smaller. We expect this family to dominate against the rest when $m=\ell=0$ and the BH is near extremal (see also Tables \ref{table} and \ref{table2}). 

It is important to note that all QNMs computed in the subextremal parameter space of the charged $C-$metric have $\text{Im}(\omega)<0$, which indicates that such spacetime is modally stable against neutral massless scalar perturbations. This result is further justified in the next section. Furthermore, we have meticulously scanned $f(r)$ and $V_r$ in the subextremal parameter space of the charged $C-$metric and found no parameters for which these functions present negative regions. Since $f(r)$ and $V_r$ are positive definite functions in the considered region, one can infer \cite{Horowitz_2000} that, indeed, solutions with $\text{Im}(\omega)\ge 0$ should not exist in the charged $C-$metric.
\begin{table}[h]
\centering
\scalebox{0.8}{
\begin{tabular}{||c| c | c ||} 
\hline
  \multicolumn{3}{||c||}{$Q/M=0.3$} \\
   \hline
    $\ell $ & $\alpha M=0.05$ & $\alpha M=0.1$  \\ [0.5ex] 
  \hline
    &$\omega_\alpha$=-0.0506 i  &$\omega_\text{PS}$=0.1079 - 0.1012 i   \\
  0 & $\omega_\text{PS}$=0.1112 - 0.1042 i &$\omega_\alpha$=-0.1055 i\\ 
   & $\lambda$=0.3317&$\lambda$=0.3269\\ 
   \hline
     & $\omega_\text{PS}$=0.2941 - 0.0976 i &$\omega_\text{PS}$=0.2839 - 0.0956 i   \\
    1  &$\omega_\alpha$=-0.1003 i  &$\omega_\alpha$=-0.2019 i\\ 
      & $\lambda$=2.3246&$\lambda$=2.2981\\ 
   \hline
    &$\omega_\text{PS}$=0.4852 - 0.0968 i & $\omega_\text{PS}$=0.4674 - 0.0954 i \\
    2  &$\omega_\alpha$=-0.1502 i   &$\omega_\alpha$=-0.3014 i\\  
      & $\lambda$=6.3100&$\lambda$=6.2392\\
   \hline
\end{tabular}}
\scalebox{0.8}{
\begin{tabular}{||c| c | c ||} 
\hline
  \multicolumn{3}{||c||}{$Q/M=0.999$} \\
   \hline
    $\ell $ & $\alpha M=0.3$ & $\alpha M=0.5$  \\ [0.5ex] 
  \hline
&$\omega_\text{NE}$=-0.0412 i  & $\omega_\text{NE}$=-0.0342 i \\ 
  0 & $\omega_\text{PS}$=0.1117 - 0.0814 i & $\omega_\text{PS}$=0.0731 - 0.0581 i \\
   &$\lambda$=0.3033  & $\lambda$=0.2497\\ 
   \hline
     & $\omega_\text{PS}$=0.3018 - 0.0782 i  &$\omega_\text{PS}$=0.1938 - 0.0553 i  \\
    1  &$\omega_\text{NE}$=-0.0819 i&$\omega_\text{NE}$=-0.0676 i\\ 
      & $\lambda$=2.1230&$\lambda$=1.7488\\ 
   \hline
   & $\omega_\text{PS}$=0.4987 - 0.0780 i &$\omega_\text{PS}$=0.3196 - 0.0550 i  \\
    2  & $\omega_\text{NE}$=-0.1228 i &$\omega_\text{NE}$=-0.1013 i\\  
      & $\lambda$=5.7626&$\lambda$=4.7468\\
   \hline
\end{tabular}}
\caption{Scalar QNMs with $m_0=0$ of the charged $C-$metric for various parameters, where $\lambda$ are the separation constants associated with the angular quantum numbers $\ell$, while $\omega_\text{PS}$ are the photon surface QNMs, $\omega_\alpha$ the acceleration modes and $\omega_\text{NE}$ the near extremal modes. The absence of acceleration modes for large $\alpha M$ is due to the fact that they are too subdominant to be captured by our numerics. The same holds for the absence of near extremal modes for small $Q/M$.}
\label{table2}
\end{table}

\begin{figure*}[t]
\subfigure{\includegraphics[scale=0.24]{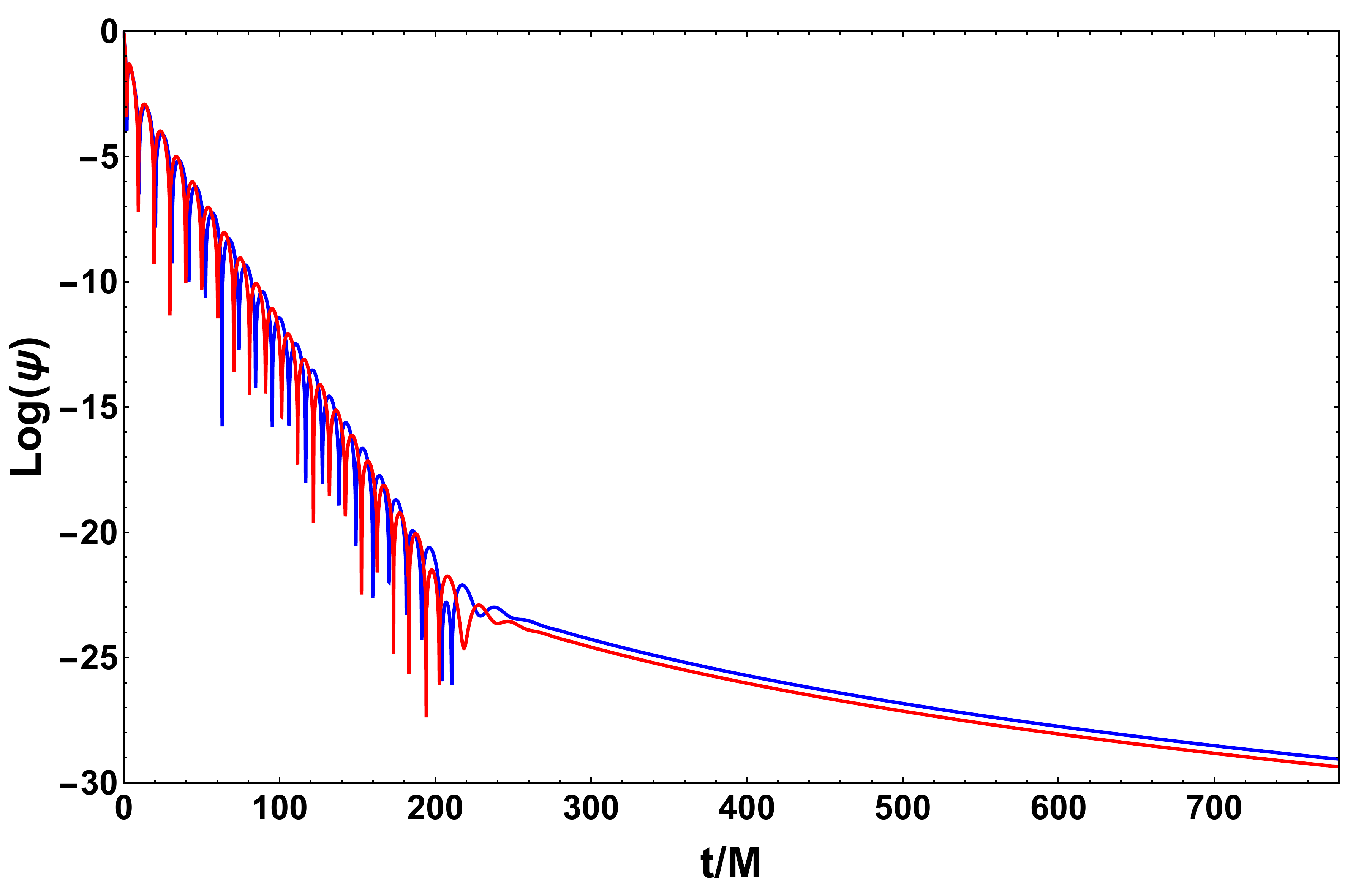}}\hskip 2ex
\subfigure{\includegraphics[scale=0.24]{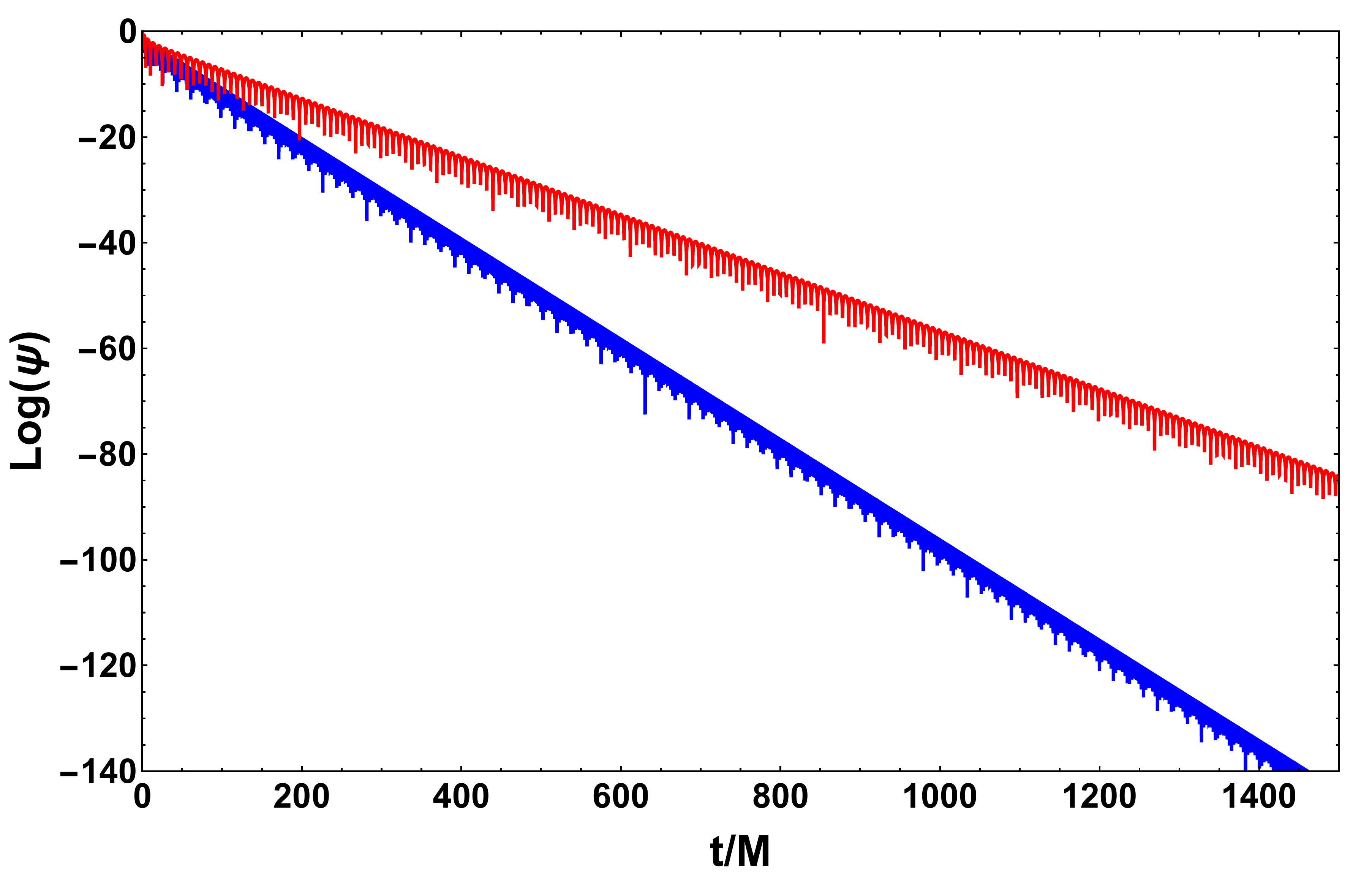}}
\subfigure{\includegraphics[scale=0.24]{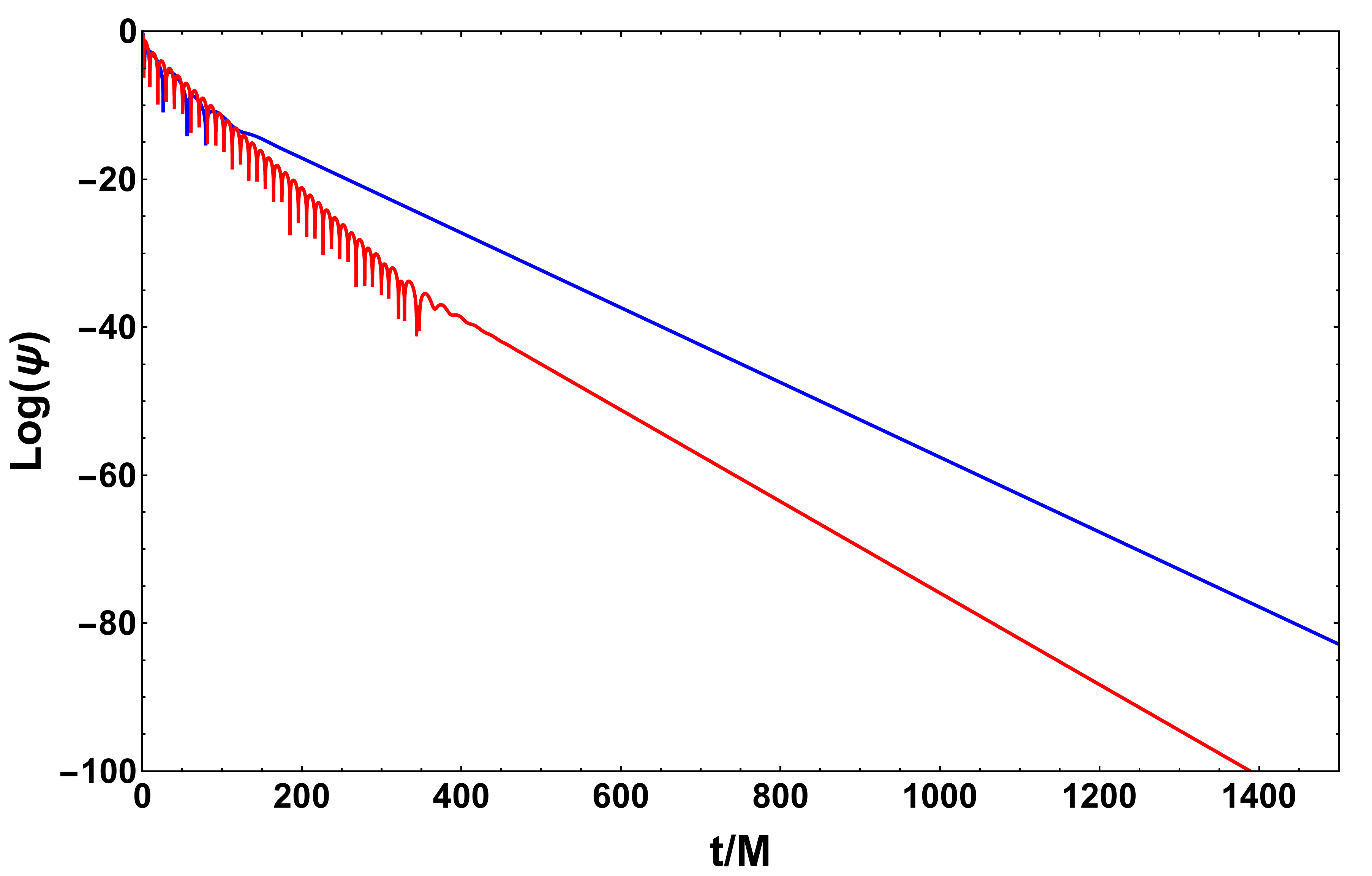}}\hskip 2ex
\subfigure{\includegraphics[scale=0.24]{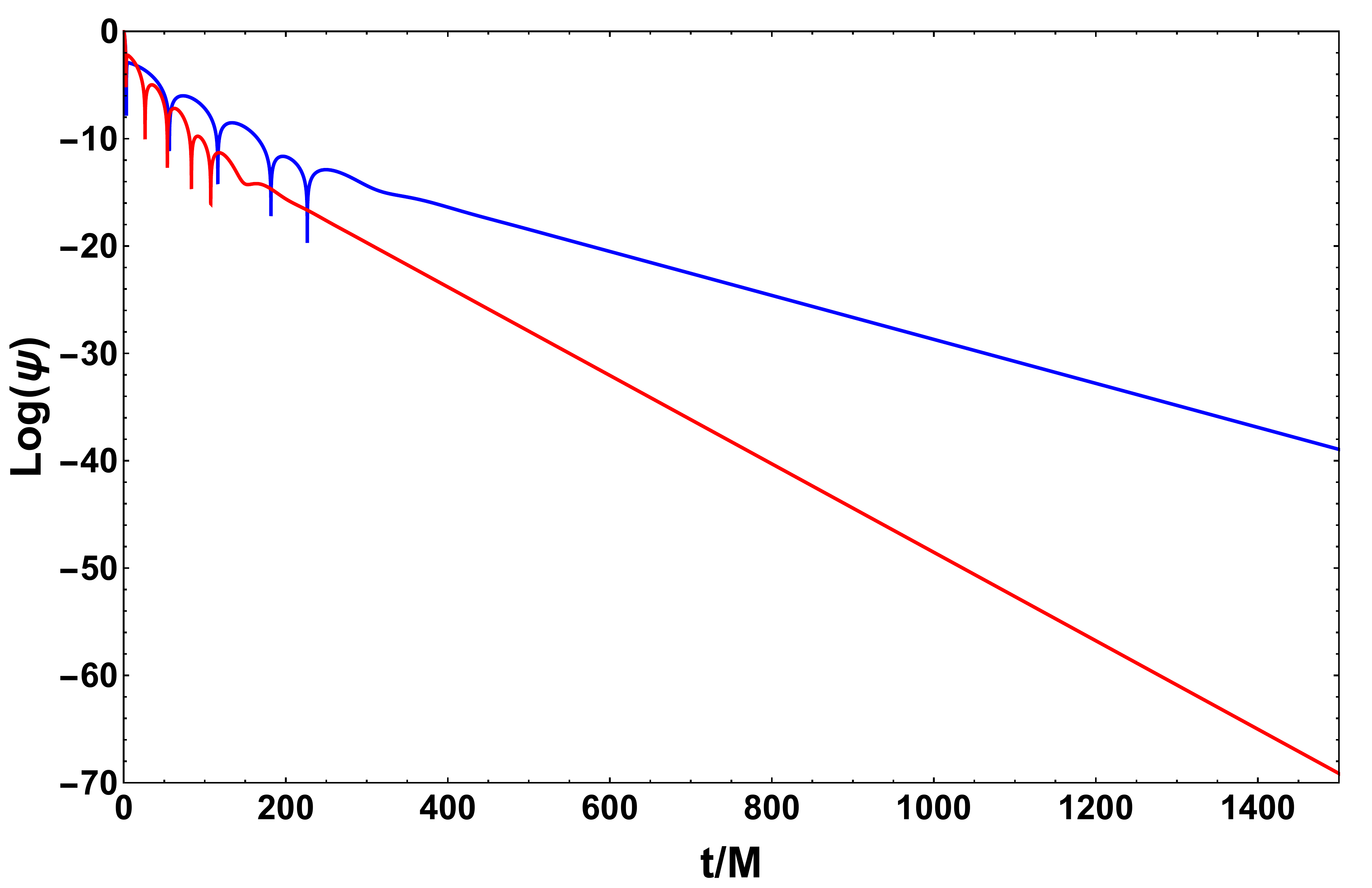}}
\caption{Top left: Time evolution of a scalar field with $\ell=1$ on a Schwarzschild (blue curve) and Reissner-Nordtr\"om (red curve) black hole with $Q/M=0.5$. Top right: Time evolution of a scalar field with $m_0=10$ on an accelerating black hole with $Q/M=0.3$, $\alpha M=0.1$ (blue curve) and a scalar field with $m_0=0$ on an accelerating black hole with $Q/M=0.999$, $\alpha M=0.5$ (red curve). The dominant QNMs in both cases belong to the photon surface family. Bottom left: Time evolution of a scalar field with $m_0=0$ on an accelerating black hole with $Q/M=0.3$, $\alpha M=0.05$ (blue curve) and a scalar field with $m_0=1$ on an accelerating black hole with $Q=0$ and $\alpha M=0.03$ (red curve). The dominant QNMs in both cases belong to the acceleration family.  Bottom right: Time evolution of a scalar field with $m_0=0$ on an accelerating black hole with $Q/M=0.9995$, $\alpha M=0.6$ (blue curve) and a scalar field with $m_0=0$ on an accelerating black hole with $Q/M=0.999$, $\alpha M=0.3$ (red curve). The dominant QNMs in both cases belong to the near extremal family. The QNMs for all cases are given in Table \ref{table_comp}.}
\label{time_evol}
\end{figure*}
%%%%%%%%%%%%%%%%%%%%%%%%%%%%%%%%%%%%%%%
\section{Stability and Late-time tails}
In this section we study the response of accelerating BHs against linear scalar perturbations in the time-domain and describe how each family of modes affects the late-time behavior of the perturbation field.

Past studies show that asymptotically flat BHs respond to external perturbation in a very distinctive way, undergoing three different stages: Initially, an outburst of radiation occurs which carries away energy through gravitational wave emission. Later, the perturbations evolve as damped oscillations (QNMs) characteristic of the BH. This stage is referred to as quasinormal ringing and does not depend on the initial configuration of the perturbation field. In the final stage, the quasinormal ringing gives way to an inverse power-law tail, first described in \cite{Price_1972,Price_1972b,Leaver}. The most complete and mathematically rigorous picture of perturbations in asymptotically flat spacetimes, to date, was provided in a series of studies \cite{Gundlach:1993tp,Gundlach_1994,Dafermos:2010hb,Dafermos:2014cua}. In turn, the behavior of radiation at fixed distances from the BH, at future null infinity and at the event horizon was described in \cite{Gundlach:1993tp,Gundlach_1994}, where their novel numerical simulations for the collapse of a self-gravitating scalar field showed that inverse power-law tails are a general feature of radiative decay, even if a BH does not form. A definitive mathematical proof of boundedness and decay results for the wave equation in Kerr spacetime, for the general subextremal case was given in \cite{Dafermos:2010hb,Dafermos:2014cua}. Thus, if the response of the perturbation field is observed at fixed radius and the field is static prior to collapse, then at $t\to\infty$
\begin{equation}
\label{power law}
\psi\sim t^{-(2\ell+2)},
\end{equation}
where $\ell$ is the angular quantum number of the perturbation.

Since the $C-$metric has no cosmological constant and is asymptotically flat (in the sense of \cite{ashtekar1981}), the behavior of the scalar field at late times could be expected to follow \eqref{power law}. Nevertheless, acceleration leads to the existence of regions where events will never intersect the world line of the accelerating black hole and thus to the formation of an acceleration horizon. So, there exists a clear physical boundary beyond which the numerical integration of the wave equation is pointless. Therefore, this boundary is where we apply our ``purely outgoing" conditions \eqref{bcs_radial}. The change of boundary conditions, and therefore the restriction of scalar field evolution between two horizons, is expected to modify the field's response significantly.

In Fig. \ref{time_evol} we present the response of a scalar field, simulated by a Gaussian wave, when propagating on a Schwarzschild (or RN) BH and on the $C-$metric. We can clearly see that when the BHs are accelerating, the late-time behavior of the scalar perturbations changes dramatically due to the modification of the boundary conditions. 

First of all, the decay is not following an inverse power-law but rather an exponential law. Secondly, and most importantly, the late-time behavior is governed by the dominant scalar QNM. More specifically, when the PS modes are the longest-lived, the late-time response does not exhibit a tail, but rather the exponentially-decaying quasinormal ringing phase is prolonged indefinitely. This is expected since the dominant PS modes are complex. When the acceleration or the NE modes are the dominant ones, an exponential tail suppresses the quasinormal ringing phase. This can be understood from the fact those families of modes are purely imaginary, therefore they do not have an oscillatory frequency.

In all cases presented in Fig. \ref{time_evol} (besides the top left where $\alpha=0$), the dominant QNM found from the frequency-domain analysis matches the QNM extracted from the waveform at late times. 

Here, we provide numerical evidence which indicate that the late-time response of the scalar field, in our accelerating BH backgrounds, indeed follows an exponential power-law
\begin{equation}
\label{expo}
\psi\sim e^{-\gamma t},
\end{equation}
where
\begin{equation}
\gamma\equiv\text{inf}_{mn}\{-\text{Im}(\omega)\}
\end{equation}
is the smallest (in absolute value) imaginary part of all families of QNMs. 
A similar decay law has also been found in asymptotically de Sitter spacetimes \cite{Hintz:2016gwb,Hintz:2016jak}. This may not be a surprise, since the cosmological horizon of de Sitter shares similar characteristics to the acceleration horizon of the boosted BHs studied here. An important insight that we can derive from such results is that the asymptotic structure of the background BH spacetime does not always dictate how the perturbation field behaves at late times.

\section{Conclusions}

Black holes have been proven to be of paramount importance to our understanding of strong-field gravity, serving as cosmic laboratories. Many black holes, throughout their lifetime, will interact with surrounding matter, absorb stars and even collide with other ultra compact objects leading to a highly perturbed state, where the black holes undergo damped oscillations until they reach a final stable state. These damped oscillations are characterized by quasinormal modes, which portray the final object and describe its externally observable properties. Therefore, the understanding of quasinormal modes, as well as the late-time behavior of the ringdown, is central to black hole physics, gravitational-wave astronomy and the asymptotic structure of spacetime.

Although the quasinormal modes of static  black holes have been abundantly investigated, there was hardly any knowledge about how ``moving" black holes vibrate. Black holes indeed move and accelerate under the gravitational influence of other compact objects, leading to binary mergers. As the objects come closer together they accelerate, due to the emission of gravitational radiation and, in certain cases, may lead to a moving remnant with velocity high enough to escape the gravitational pull of its host galaxy.
 
A natural choice to describe accelerating black holes is the $C-$metric. This metric describes two black holes accelerating away from each other in opposite directions, under the influence of a cosmic string in tension. By a suitable choice of coordinates, one can focus on one of the accelerating black holes, where the acceleration is encoded through a boost parameter in the metric tensor.

In this article, we have calculated numerically the scalar quasinormal modes of the $C-$metric and its charged version, for the first time, by using a frequency and a time-domain analysis. We have identified three distinct families of quasinormal modes which do not converge to one another in any limit. 

The first family is associated to the photon surface of the $C-$metric, is complex, and undergoes an anomalous behavior under the increment of the boost parameter. For a sufficiently small boost, the longest-lived modes are the ones with large frequencies, while beyond a threshold of the boost parameter the longest-lived modes are those with small frequencies. This anomalous effect is significantly suppressed when the accelerating black hole is charged. 

The second family of modes is purely imaginary and is completely characterized by the existence of an acceleration horizon. These modes seem robust to the addition of electric charge to the black hole and only depend on the acceleration parameter. As far as we are aware, this family had not been discovered before.

The final family is related to the event horizon temperature of the near extremally charged $C-$metric and becomes long-lived when the event and Cauchy horizons are about to meet. The increment of the acceleration parameter makes these modes decay even slower. 

The time-domain analysis of scalar perturbations in accelerating black holes reveals a late-time behavior which, to our knowledge, has never been observed before in black holes without a cosmological constant. In contrast to the anticipated power-law cutoff of perturbations, which non-accelerating asymptotically flat black holes exhibit, accelerating black holes endure an infinitely-prolonged exponential ringdown which is characterized by the dominant quasinormal modes. We identify this outcome with the presence, and the nature, of the acceleration horizon, which imposes a change of boundary conditions. This result suggests that the asymptotic structure of background black hole spacetimes does not always dictate how radiative fields behave asymptotically in time.

Here, we have narrowed our study to neutral scalar fields on non-rotating accelerating black holes. Nevertheless, rotating accelerating black holes do exist, as exact solutions to the field equations, as well as accelerating black holes with a NUT charge in asymptotically de Sitter and anti-de Sitter spacetimes. The extension of our results in such directions, as well as the inclusion of charge in the perturbing scalar field, would be very interesting if one considers the similarities shared between the acceleration and cosmological horizon of de Sitter and the existence of superradiance, and superradiant instabilities, of charged scalar fields in Reissner-Nordstr\"om-(anti)de Sitter and Kerr-Newman-(anti)de Sitter black holes \cite{Brito:2015oca}.

Furthermore, the geometrical similarities of acceleration and de Sitter horizons can be exploited to study the extendibility of solutions beyond the Cauchy horizon of the charged $C-$metric and the validity of the modern formulation of the Strong Cosmic Censorship conjecture (see e.g. \cite{Cardoso:2017soq}). Accelerating black holes obey all the criteria needed to test the linear stability of Cauchy horizons, namely the exponential decay of scalar perturbations dominated by quasinormal modes and the exponential blueshift mechanism triggered by such perturbations in the black hole interior. Since both mechanisms are exponential, they can antagonize and, in some regime, counterbalance each other, leading to a stable enough Cauchy horizon which may allow observers to cross beyond it smoothly. We address this question on a follow up paper \cite{Destounis:2020yav}.
\\~\\
{\noindent \bf{Acknowledgements.}} The authors are grateful to Vitor Cardoso, David Kofroň and Rodrigo Vicente for helpful discussions. KD acknowledges financial support and hospitality from CAMGSD, IST, Univ. Lisboa where this project was initiated. KD also acknowledges networking support by the GWverse COST Action CA16104, "Black holes, gravitational waves and fundamental physics". FCM thanks support from CAMGSD, IST, Univ. Lisboa, and CMAT, Univ. Minho, through FCT funds UID/MAT/04459/2019 and Est-OE/MAT/UIDB/00013/2020, respectively, and FEDER Funds COMPETE.
\appendix
\section{Comparison of methods}\label{appa}
In this appendix, we show results obtained through our time and frequency-domain QNM calculations. Table \ref{table_comp} shows the agreement between our numerical methods.
\\~\\
\begin{table}[h]
\centering
\scalebox{0.63}{
\begin{tabular}{||c | c| c | c ||} 
\hline
    $\text{Case}$&n & $\text{time domain}$ & $\text{frequency domain}$  \\ [0.5ex] 
  \hline
   $Q/M=0.3$, &0 &$\omega_\text{PS}$=2.332170 - 0.0951217 i &$\omega_\text{PS}$=2.3321170 - 0.095128 i   \\
  $\alpha M=0.1$, &1 & $\omega_\text{PS}$=2.32887 - 0.2855 i &$\omega_\text{PS}$=2.32882 - 0.28551 i\\
  $m_0=10$, &2 & $\omega_\text{PS}$=2.3223 - 0.4763 i &$\omega_\text{PS}$=2.3222 - 0.4763 i\\
  $\lambda$=155.4315 & & &\\ 
   \hline
   $Q/M=0.999$, &0 & $\omega_\text{PS}$=0.401597 - 0.054934 i &$\omega_\text{PS}$=0.401596 - 0.054935 i   \\
     $\alpha M=0.5$,&1 &$\omega_\text{NE}$=-0.1231669 i  &$\omega_\text{NE}$=-0.1231669 i\\ 
    $m_0=1$, &2 &$\omega_\text{NE}$=-0.157482 i  &$\omega_\text{NE}$=-0.157483 i\\ 
     $\lambda$=7.4893 && &\\ 
   \hline
   $Q/M=0.3$, &0&$\omega_\alpha$=-0.0505665 i &$\omega_\alpha$=-0.0505665 i  \\
    $\alpha M=0.05$,&1  & $\omega_\alpha$=-0.10344 i  &$\omega_\alpha$=-0.10344 i\\ 
    $m_0=0$, &2&$\omega_\text{PS}$=0.11117 - 0.104246 i   &$\omega_\text{PS}$=0.11117 - 0.104247 i\\  
    $\lambda$=0.3317  & & &\\
    \hline
     $Q/M=0$,  &0&$\omega_\alpha$=-0.06189606 i& $\omega_\alpha$=-0.06189607 i \\
       $\alpha M=0.03$,& 1 &  $\omega_\alpha$=-0.0921 i  &$\omega_\alpha$=-0.0921 i\\ 
       $m_0=1$, &2&$\omega_\text{PS}$=0.30325 - 0.097363 i   &$\omega_\text{PS}$=0.30325 - 0.097367 i\\  
        $\lambda$=2.5177 & & &\\   
   \hline
    $Q/M=0.9995$,&0&$\omega_\text{NE}$=-0.02049096 i & $\omega_\text{NE}$=-0.02049096 i \\
          $\alpha M=0.6$,& 1 &$\omega_\text{NE}$=-0.041720 i   &$\omega_\text{NE}$=-0.041720 i\\ 
          $m_0=0$, &2&$\omega_\text{PS}$=0.0517195 - 0.0429256 i   &$\omega_\text{PS}$=0.0517195 - 0.0429256 i\\  
          $\lambda$=0.2131  && &\\   
      \hline
  $Q/M=0.999$, &0&$\omega_\text{NE}$=-0.04120898 i & $\omega_\text{NE}$=-0.04120898 i \\
             $\alpha M=0.3$, &1 &$\omega_\text{PS}$=0.1116755 - 0.0814151 i   &$\omega_\text{PS}$=0.1116754 - 0.0814153 i\\ 
             $m_0=0$, &2&$\omega_\text{NE}$=-0.0836409 i   &$\omega_\text{NE}$=-0.0836409 i\\  
             $\lambda$=0.3033  & &&\\   
         \hline   
\end{tabular}}
\caption{Comparison between the time and frequency domain integration methods used for various cases. We recall that $\lambda$ represents the separation constants associated with the angular quantum numbers $\ell$, while $\omega_\text{PS}$ are the photon surface QNMs, $\omega_\alpha$ the acceleration modes and $\omega_\text{NE}$ the near extremal modes. }
\label{table_comp}
\end{table}
\FloatBarrier

%%%%%%%%%%%%%%%%%%%%%%%%%%%%%%%%%%%%%%%%%%%%%%%%%%%%
\section{Frobenius method for equation (\ref{final_polar})}\label{appb}

In this appendix we describe an adaptation of the Frobenius method which we use to solve equation (\ref{final_polar}), and compare with the results from the methods mentioned at the end of Section \ref{swe}.

We begin by changing the angular coordinate with the substitution,
\be
x=\frac{1-\cos \theta}{2},
\ee
after which the angular equation (\ref{final_polar}) takes the form
\begin{equation}
\label{angular-eq}
T(T\chi')' - \frac{2}{3\alpha^2 Q^2}(6m^2+\alpha^2 Q^2 T(6\lambda + T''))\chi=0,
\end{equation}
where $T(x):=16\prod_{i=1}^4(x-x_i)$ is a polynomial with the four regular singular points of equation \eqref{angular-eq}:
\be
x_1=0, \hspace{0.5cm} x_2=1, \hspace{0.5cm} x_{3,4} = \frac{\alpha Q^2 - r_\pm}{2\alpha Q^2}.
\ee
In agreement with the boundary conditions, we choose as ansatz
\be
\label{asz}
\chi (x) = (x-1)^\gamma \sum_{n=0}^{\nu} a_n x^{\delta + n},
\ee
which upon substitution in \eqref{angular-eq} yields the indicial relation
\be
\delta  = \frac{m}{2(1-2\alpha M+\alpha^2 Q^2)}.
\ee
Now, the recurrence relation that comes from the above ansatz is given by
\be
a_n=\Delta_n\sum_{i=0}^{n-1}\Big((v_{2+i}+\tau_{2+i})(i+1-n-\delta) - (t_{1+i}
 \nonumber
\\
\nonumber
+w_{1+i}+u_{1+i})+s_{3+i}(-\delta^2 + \delta (3+2i-2n) \\
-(n-1-i)(n-2-i)\Big)a_{n-i-1},\hspace{1cm}
\ee
 in which 
\be
\nonumber
\Delta_n=\frac{1}{u_0 + \tau_1(\delta + n) + s_2 (\delta^2 + \delta (2n-1)+n^2-n)},
\ee
and the terms $v_n,\tau_n , t_n, w_n, u_n$ and $s_n$ are the expansion coefficients of the equation with the ansatz (\ref{asz}), written as
\be
\nonumber
w & = & \gamma (\gamma - 1) (x-x_2)^{-2}T^2 \equiv \sum_{n=0}^{\nu} w_n x^n, \\
\nonumber
v & = & 2 \gamma (x-x_2)^{-1}T^2 \equiv \sum_{n=0}^{\nu} v_n x^n, \\
\nonumber
s & = & T^2 \equiv \sum_{n=0}^{\nu} s_n x^n, \\
\nonumber
t & = & \gamma (x-x_2)^{-1}TT' \equiv \sum_{n=0}^{\nu} t_n x^n,\\
\nonumber
\tau & = & TT' \equiv \sum_{n=0}^{\nu} \tau_n x^n, \\
\nonumber
u & = & -4 \left( m^2+\alpha^2 Q^2 \left( \lambda + \frac{T''}{6}\right) \right)  \equiv \sum_{n=0}^{\nu} u_n x^n.
\ee
In turn, the second boundary condition leads to 
\be
\chi\Big|_{x\rightarrow 0} = a_0x^\delta,
\ee
which remains physical as long as $a_0$ is bounded. For the purpose our problem, we can freely choose $a_0=1$. Regarding the first boundary condition, we have two relations to be specified, namely
\be
\gamma = \frac{m_0}{2},
\ee
where we recall that $m=m_0 P(\pi)$, and 
\be
 \chi\Big|_{x\rightarrow 1} = \sum_{n=0}^{\nu}a_n(x-1)^{\gamma},
\ee
which remains physical as long as the sum is bounded. Then, the eigenvalue $\lambda$ comes as a solution of the system 
\be
\sum_{n=0}^{\nu}a_n = C_2,
\ee
with $a_0 = C_1$, where $C_1$ and $C_2$ are constants. The radius of convergence $R$ of the ansatz, spans from the point of expansion $x_1$ to the next singular point, in general $x_2$, so $R=1$. In this way, $C_2$ is marginally convergent and no matter how far we increase $\nu$, the computational effort to determine the eigenvalue will fail. For instance, taking $m_0=0$ and small enough values for $Q$ and $\alpha$ (i.e. the most convergent of all cases), we need at least $\nu = 15000$ to reproduce the first eigenvalue with an accuracy of $3\% $\footnote{In any usual 'i5' computer this takes nearly one day of computational time.}. In order to avoid this convergence difficulty, we propose another approximation to the problem based on the fact that every $\lambda$, which is actually an eigenvalue, is the best locally convergent $C_2$ (i.e. a minimum). In terms of the computational implementation, this represents a ``bissectional-inspired" method. The first step is the choice of a specific $\nu$, followed by a test function 
\be
\label{minf}
f(\lambda_i)= \Bigg|1 - \frac{\sum_{n=0}^{\nu}a_n}{\sum_{n=0}^{\nu /2}a_n} \Bigg|
\ee
calculated for different $\lambda_i= k+i\alpha$, $i=1,2,3,\cdots,i_f$. The essence of the method is to find a minimum for $f(\lambda_i)$ in a given iteration and refine the grid parameter $\alpha$ and the starting $\lambda$-shooting value $k$ for the next iteration. The approximative guess for $\lambda$ is to be taken not far away from $l(l+1) + 1/3$ such that the choices of both $k$ and $\alpha$ are bounded. 

As an example, in Table \ref{eigen} we show the convergence of the method for 19 steps of iteration for the parameters $M=5Q=25\alpha = m_0 = 1$ with $\nu = 500$.
\begin{table}[h]
\centering
\scalebox{0.63}{
\begin{tabular}{|| c | c | c | c | c ||} 
\hline
 Iteration & $k$ & $\alpha$ & $i_f$ & $\lambda_m$  \\
  \hline
      $1$ &  $0$ & $0.4$ & $10$ & $2.4$\\
   \hline
      $2$ &  $2.0$ & $0.1$ & $10$ & $2.5$\\
   \hline
      $3$ &  $2.4$ & $0.02$ & $10$ & $2.58$\\
   \hline 
      $4$ &  $2.56$ & $0.004$ & $10$ & $2.580$\\
   \hline
      $5$ &  $2.576$ & $0.001$ & $10$ & $2.581$\\
   \hline 
      $6$ &  $2.580$ & $0.0002$ & $10$ & $2.5812$\\
   \hline
      $7$ &  $2.5810$ & $0.00004$ & $10$ & $2.58124$\\
   \hline
      $8$ &  $2.58120$ & $0.00001$ & $10$ & $2.58126$\\
   \hline
      $\cdots$ &  $\cdots$ & $\cdots$ & $\cdots$ & $\cdots$\\ 
   \hline
      $19$ &  $2.581255950394$ & $4.10^{-13}$ & $10$ & $2.5812559503968$\\ 
   \hline
\end{tabular}}
\caption{Convergence of $\lambda$ for the Frobenius method with the function (\ref{minf}). Here $\lambda_m$ stands for the value of $\lambda$ in each iteration for which $f$ is minimum.}
\label{eigen}
\end{table}
With the above approximation, the correct $\lambda$ eigenvalue stands for the last $\lambda_m$ ($19^{th}$ iteration) and corresponds exactly to the same value calculated with the method of spectral decomposition used in Section \ref{swe}.

The method is liable when $x_3$ and $x_4$ are not within the convergence radius of our expansion, $R=1$, resulting in the convergence condition 
\be
3 \alpha r_+ < 1.
\ee

\bibliography{references}
\end{document}